\documentclass[sigconf]{acmart}

\setcopyright{none}
\copyrightyear{2027}
\acmYear{2027}
\acmISBN{}
\acmDOI{}
\acmConference[KDD '27]{Proceedings of the 33rd ACM SIGKDD Conference on Knowledge Discovery and Data Mining}{2027}{Datasets and Benchmarks Track}

\usepackage{booktabs}
\usepackage{enumitem}
\usepackage{amsmath}
\usepackage{pifont}
\usepackage{graphicx}
\usepackage{capt-of}
\usepackage{longtable}
\usepackage{float}
\graphicspath{{figures/}}

\begin{document}

\title[TIDE: A Physically Diverse 3D Turbulence Benchmark Dataset]{TIDE: A Physically Diverse 3D Turbulence Benchmark Dataset \\for Advancing Scientific Machine Learning}

\author{Yilong Dai}
\affiliation{%
  \institution{University of Alabama}
  \city{Tuscaloosa}
  \country{USA}}
\email{ydai17@ua.edu}

\author{Yiming Sun}
\affiliation{%
  \institution{University of Pittsburgh}
  \city{Pittsburgh}
  \country{USA}}
\email{yimingsun@pitt.edu}

\author{Yiheng Chen}
\affiliation{%
  \institution{University of Alabama}
  \city{Tuscaloosa}
  \country{USA}}
\email{ychen226@ua.edu}

\author{Shengyu Chen}
\affiliation{%
  \institution{University of Pittsburgh}
  \city{Pittsburgh}
  \country{USA}}
\email{shc160@pitt.edu}

\author{Peyman Givi}
\affiliation{%
  \institution{University of Pittsburgh}
  \city{Pittsburgh}
  \country{USA}}
\email{peg10@pitt.edu}

\author{Xiaowei Jia}
\affiliation{%
  \institution{University of Pittsburgh}
  \city{Pittsburgh}
  \country{USA}}
\email{xiaowei@pitt.edu}

\author{Runlong Yu}
\authornote{Corresponding author.}
\affiliation{%
  \institution{University of Alabama}
  \city{Tuscaloosa}
  \country{USA}}
\email{ryu5@ua.edu}

\begin{abstract}
Turbulence is a central testbed for machine learning on physical dynamics because its governing laws are known exactly. However, most existing studies remain in 2D, while 3D turbulence has fundamentally different physics and is far more costly to simulate. Existing 3D resources also typically provide only one realization per configuration, making it difficult to distinguish learning the dynamics from fitting the statistics of a single flow. In this paper, we introduce \textbf{TIDE} (\textbf{T}urbulent \textbf{I}ncompressible \textbf{D}NS \textbf{E}nsembles), a $256^3$ DNS corpus and benchmark for 3D incompressible turbulence, with 15 configurations on eight controlled axes, independent ensembles, pressure fields, and equation-level verification. The benchmark includes five tasks, standardized learned baselines, controlled generalization splits, and physical-fidelity metrics alongside pointwise error. Across the main forecasting configurations, current learned models trained under a fixed reference recipe barely outperform persistence and still make about twice the error of a spectral solver given the true equations. Moreover, lower pointwise error can coincide with severely distorted small-scale dynamics, showing that accuracy alone does not ensure physical fidelity. Rollout stability depends on the configuration, the training seed, and how densely training samples each trajectory. Joint training across neighboring regimes does not close the transfer gaps we test, and the forced-to-decay contrast bounds a conditioning limit: an estimate of the current forcing state removes 20--41\% of the error a spectral solver makes without forcing input, while history or forcing inputs help only some learned architectures. Closing these accuracy, fidelity, and conditioning gaps is the central open problem made measurable by TIDE. https://github.com/Dyloong1/TIDE-dataset-benchmark
\end{abstract}


\maketitle

\section{Introduction}
\label{sec:intro}

Turbulence is among the oldest unsolved problems of classical physics and
among the most consequential, setting the drag, mixing, and dissipation
of most flows of engineering and geophysical
interest~\cite{pope2000turbulent}. Its canonical setting is a periodic
box, where direct numerical simulation (DNS) resolves all dynamically
active scales with no turbulence model. For machine learning this box
has become a standard proving ground: the governing laws are known
exactly, the data are the archetype of high-dimensional multiscale
chaos, and a large body of methods is developed and validated on
turbulent
flows~\cite{raissi2019pinn,li2021fno,lu2021deeponet,wang2020tfnet,ling2016reynolds,duraisamy2019turbulence,kochkov2021ml,um2020solver,fukami2019super,kohl2024acdm,lippe2023pderefiner,stachenfeld2022learned,brunton2020ml,dai2026flowlearnerspdesphysicstophysics}.
Methods and evaluation practices developed on this testbed carry over to
data-driven mechanics more broadly. Most of this work, however, reports
results in 2D or at modest 3D resolution.

\begin{figure}[t]
  \centering
  \includegraphics[width=0.85\linewidth]{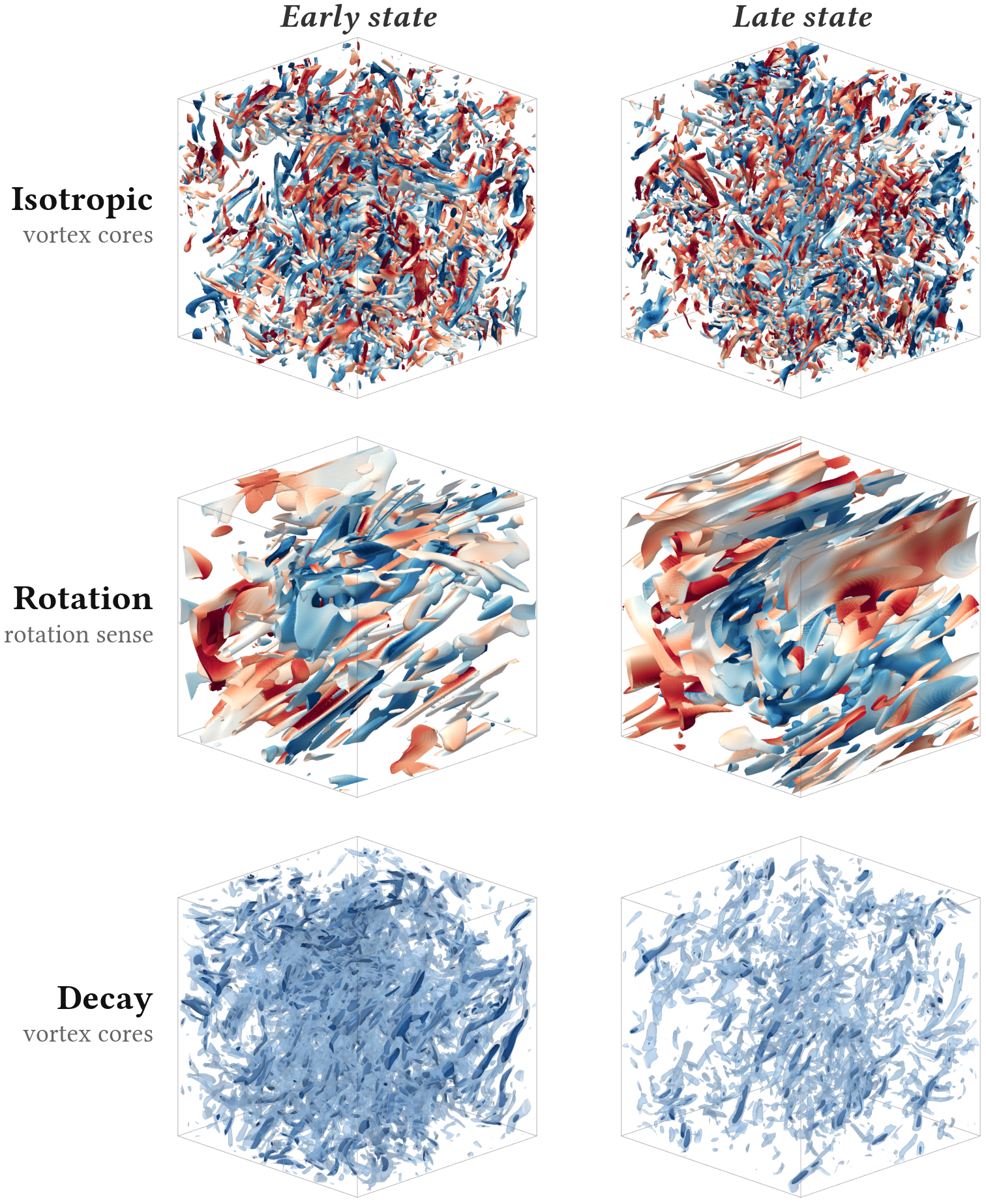}
  \caption{Vortex structure of three regimes at an early and a late
  instant: $Q$-criterion isosurfaces colored by axial vorticity
  (isotropic, rotation); vorticity isosurfaces (decay).}
  \label{fig:hero3d}
\end{figure}

\begin{figure}[t]
  \centering
  \includegraphics[width=\linewidth]{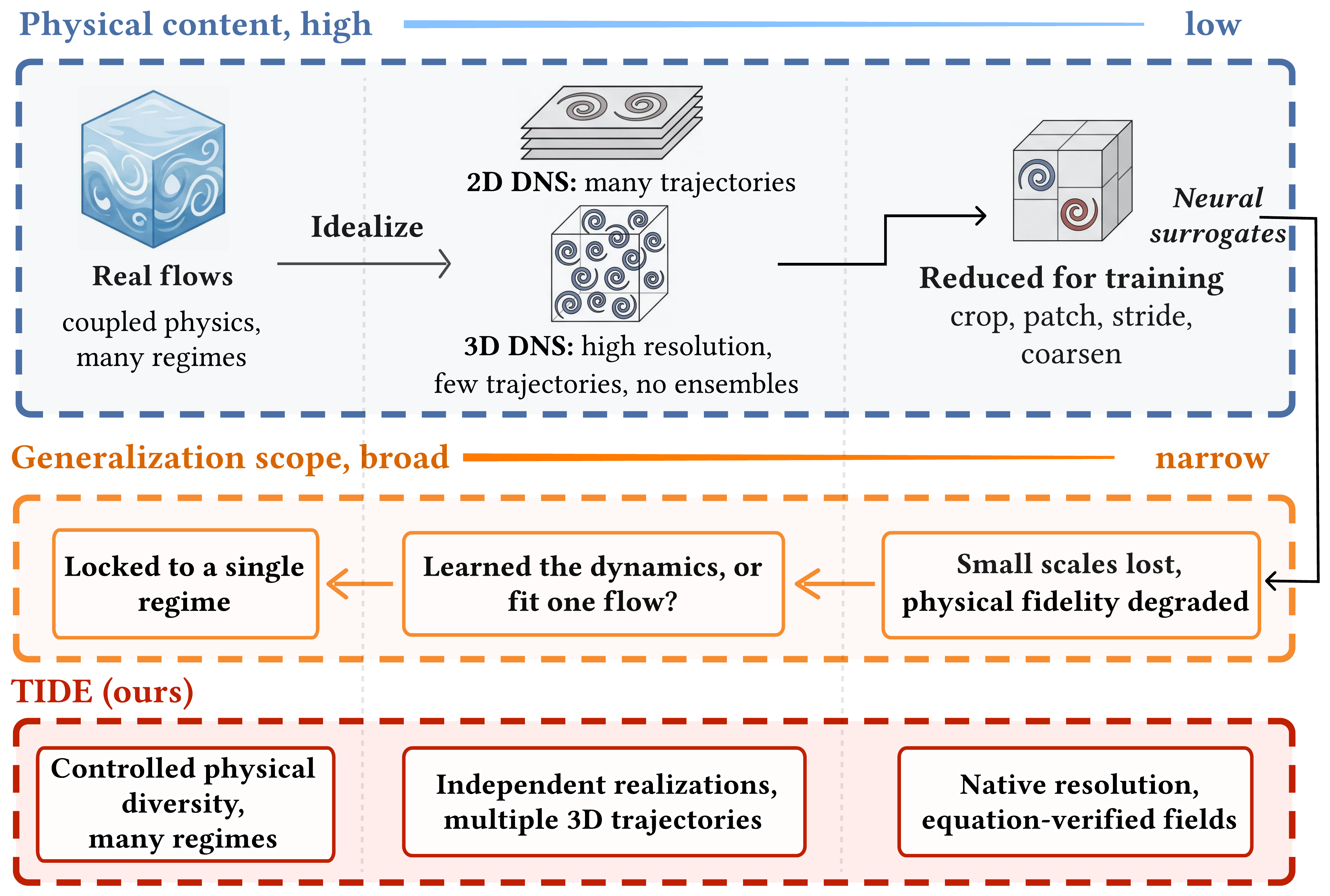}
  \caption{Overview of the machine-learning pipeline for turbulence: the
  loss of physical content from real flows to training inputs (top), the
  resulting limits in generalization and dynamical identification
  (middle), and TIDE's corresponding design choices (bottom).}
  \vspace{-0.2cm}
  \label{fig:pipeline}
\end{figure}

That concentration matters, because the central difficulty of turbulence
is specifically three-dimensional: vortex stretching, the mechanism that
drives energy toward small scales, exists only in 3D, and 2D turbulence
moves energy in the opposite
direction~\cite{boffetta2012twodimensional}. Two dimensions are a different
physical system, not a lower-resolution proxy; Figure~\ref{fig:hero3d}
makes the difference visible, with vortex structures tangling in all
three dimensions. However, high-resolution 3D data are costly to
generate, with computational expense rising steeply with resolution. As
a result, general-purpose PDE collections remain dominated by 1D and 2D
fluid systems~\cite{takamoto2022pdebench,ohana2024thewell}.

High-fidelity 3D DNS is not itself what is missing: turbulence-native
databases host landmark simulations for selected turbulent flows at
higher turbulence intensities~\cite{li2008jhtdb}. What is missing is the
\emph{structure aimed at machine learning}: Figure~\ref{fig:pipeline}
traces how physical content thins along the current pipeline, and how
each failure it produces maps back to a stage of that reduction.
Existing turbulence databases typically provide one realization per
configuration~\cite{li2008jhtdb,lee2015channel,hoyas2006scaling}. Hence, a model trained and tested
within one flow can score well without settling the central question of
whether it learned the dynamics or fit the statistical signature of that
one flow. Additionally, 3D ML-curated corpora largely target different governing
systems, such as compressible reacting flow~\cite{chung2023blastnet}.

Determining whether a model has learned the underlying dynamics rather
than merely fit the statistical signature of a particular flow requires
\emph{diversity of the physical regime}: the
same equations with the energy budget changed, a symmetry broken, or the
drive removed. Constructing that contrast requires holding the solver,
the resolution, and the acceptance standard fixed while only the physics
varies. To our knowledge, no existing PDE or turbulence data source is
organized this way
(Table~\ref{tab:related}). The
regime shifts are also not equivalent: most are visible in the flow
itself, so adaptation is in principle a matter of training
coverage. In contrast, the drive itself leaves no signature on a single early
snapshot: whether a flow is driven or freshly decaying cannot be read
off one frame. Our benchmark is built around
this distinction (Section~\ref{sec:benchmark}).

The shortage of such data becomes more pronounced at the scale
required by modern machine learning. Unlike text and images, which
accumulate as a byproduct of human activity, 3D turbulence samples must
be \emph{manufactured} at DNS cost. Moreover, because the governing
equations are known exactly, turbulence samples can be checked against
those equations before being accepted, which is what our acceptance
procedure does. For 3D incompressible turbulence there is, to our
knowledge, no shared large-scale pretraining corpus and no foundation
model~\cite{mccabe2023mpp,hao2024dpot,herde2024poseidon,jiang2025diafno,datafree3dns2025}.

We address these gaps with \textbf{TIDE} (\emph{Turbulent Incompressible
DNS Ensembles}), a $256^3$ fp64 corpus built around controlled
structure: 15 configurations on eight controlled axes, each passing a
fixed acceptance standard of statistical gates and equation-level checks
before release (Section~\ref{sec:acceptance}), each shipping the
pressure channel, and each carrying 8 to 16 fully independent
realizations (134 trajectories, roughly 2.6\,TB). Independent
realizations make stability under re-realization measurable; held-out
configurations make transfer along a named physical axis measurable; and
the forced/decay contrast isolates what an operator is conditioned on,
since it removes the drive while leaving the governing equations
untouched (Section~\ref{sec:benchmark}). The axes are physically named
rather than orthogonal and the accessible turbulence intensity is
moderate; both boundaries are stated in Appendix~\ref{app:discussion}.

A companion benchmark turns the structure into measurements: five
reference tasks, five learned baselines under one fixed protocol, and
generalization scored on held-out configurations, labeled by the axis
varied. Physical-fidelity metrics are first-class alongside pointwise
error, because a single averaged error misses the small-scale departures
we measure in most rollout cells. Our contributions are summarized as
follows:
\begin{itemize}[leftmargin=1.2em,itemsep=1pt,topsep=1pt]
\item \textbf{Dataset.} To our knowledge, the first corpus of 3D
\emph{incompressible} turbulence that is both DNS-verified and
physically diverse: 15 configurations on eight controlled axes,
independent per-configuration ensembles, a forced/decay contrast, and a
pressure channel.
\item \textbf{Verification as a data property.} A unified data
validation protocol in which every configuration
satisfies a fixed acceptance standard combining statistical gates with
equation-level residuals, with the verification procedure itself
validated against known-answer fields (Section~\ref{sec:acceptance}).
\item \textbf{Benchmark.} Five tasks, five audited baselines, and
generalization splits along the controlled axes; the forced/decay split
isolates the conditioning question, which we bound with the
equations-informed integrator and test with history and forcing inputs
(Sections~\ref{sec:benchmark} and~\ref{sec:results}).
\item \textbf{Physical-fidelity evaluation.} A three-axis evaluation
framework reporting pointwise error,
small-scale physical health, and spectral fidelity,
all of which are necessary for reliable assessment
(Sections~\ref{sec:benchmark} and~\ref{sec:three-axis}).
\end{itemize}
Dataset, generation code, acceptance scripts, and the benchmark code are
publicly available.%
\footnote{Corpus: \url{https://huggingface.co/datasets/ydai17/TIDE};
DOI record: \url{https://doi.org/10.5281/zenodo.21589489}; code:
\url{https://github.com/Dyloong1/TIDE-dataset-benchmark}. Data under
CC-BY-4.0, code under MIT.}

\section{Related Work and the Position of TIDE}
\label{sec:related}

\begin{table*}[t]
\centering
\caption{TIDE vs.\ existing PDE/turbulence ML resources. Ens./cfg =
independent realizations per configuration; F/D = forced and decaying
regimes; \ding{55}\ = not described in the resource's public
documentation.}
\label{tab:related}
\small
\setlength{\tabcolsep}{3pt}
\resizebox{\linewidth}{!}{\begin{tabular}{@{}lccccccccccc@{}}
\toprule
\textbf{Resource} & \textbf{Dim} & \textbf{DNS} & \textbf{Pressure} & \textbf{Ens./cfg} & \textbf{3D turb.\ traj.} & \textbf{F/D} & \textbf{Phys.\ eval} & \textbf{Eqn.-verif.} & \textbf{Ctrl.\ OOD} & \textbf{Multi-task} & \textbf{Time-series} \\
\midrule
PDEBench \cite{takamoto2022pdebench} & 1--3D & partial & \ding{55} & \ding{55} & --- & --- & partial & --- & param.\ & \checkmark & \checkmark \\
The Well \cite{ohana2024thewell}     & 2D/few 3D & mixed & \ding{55} & \ding{55} & --- & --- & partial & --- & cross-set & partial & \checkmark \\
APEBench \cite{koehler2024apebench}  & 1--2D & spectral & \ding{55} & \ding{55} & --- & --- & spectral & --- & param.\ & rollout & \checkmark \\
SuperBench \cite{ren2023superbench}  & 2D & --- & \ding{55} & \ding{55} & --- & --- & spectral & --- & --- & \ding{55} & \ding{55} \\
CFDBench \cite{luo2023cfdbench}      & 2D & RANS/CFD & partial & \ding{55} & --- & --- & \ding{55} & --- & geom./param. & \checkmark & \checkmark \\
BLASTNet 2.0 \cite{chung2023blastnet} & 3D & \checkmark$^{c}$ & partial & \ding{55} & 744$^{c}$ & --- & spectral & --- & param.\ & super-res & partial \\
JHTDB \cite{li2008jhtdb}             & 3D & \checkmark & \checkmark & 1$^{1r}$ & 1/config & forced & DB-only & --- & n/a$^{d}$ & n/a$^{d}$ & \checkmark \\
\midrule
\textbf{TIDE (ours)} & \textbf{3D} & \textbf{\checkmark}$^{b}$ & \textbf{\checkmark} & \textbf{8--16} & \textbf{134 (15 cfg)} & \textbf{both} & \textbf{\checkmark} & \textbf{\checkmark} & \textbf{\checkmark}$^{a}$ & \textbf{\checkmark} & \textbf{\checkmark} \\
\bottomrule
\end{tabular}}
\\[2pt]
{\footnotesize $^{a}$controlled axes along which a configuration can be
held out; only $Re_\lambda$ is single-variable by construction
(Section~\ref{sec:generation}). $^{b}$all configurations at $256^3$.
$^{c}$compressible reacting flows; 744 samples over
34 conditions. $^{d}$query database; benchmark columns do not apply.
$^{1r}$one realization per dataset, per the public index.}
\vspace{-0.2cm}
\end{table*}

\paragraph{Turbulence as a testbed for learning on physical dynamics.}
Because its governing equations are known exactly, turbulence is a
standard proving ground for machine learning on physical dynamics:
physics-informed networks recover fields from sparse or indirect
observations~\cite{raissi2019pinn,raissi2020hidden}, physics-guided
architectures add divergence and boundary
constraints~\cite{wang2020tfnet}, and related lines learn
closures~\cite{ling2016reynolds,duraisamy2019turbulence} or super-resolve
coarse fields~\cite{fukami2019super,dai2026physicspreservinglatentcompressionzeroshot}, surveyed
in~\cite{brunton2020ml,dai2026learningpdesolversphysics}. The dominant surrogate families are likewise
developed and validated on turbulent flows:
FNO~\cite{li2021fno}, DeepONet~\cite{lu2021deeponet}, tensorized
variants~\cite{kossaifi2023tfno}, attention
operators~\cite{wu2024transolver,li2023factformer,dai2026pestphysicsenhancedswintransformer}, diffusion and
refinement models~\cite{kohl2024acdm,lippe2023pderefiner,dai2026flowrefinerflowmatchingbasediterative},
learned-correction solvers~\cite{kochkov2021ml}, and long-horizon rollout
studies~\cite{wu2025dino}. The size of this literature is itself the
evidence that turbulence is a first-class testbed for AI for Science,
yet its evaluations remain concentrated in 2D or on single flow
configurations: the studies reaching 3D incompressible turbulence
generate their own corpora under their own
protocols~\cite{jiang2025iafno,jiang2025diafno} or avoid training data
altogether~\cite{datafree3dns2025}, and PDE foundation models document
pretraining corpora that are predominantly
1D/2D~\cite{mccabe2023mpp,hao2024dpot,herde2024poseidon}. What this
weight of methods lacks is a shared 3D-native testbed.

\paragraph{Data generation and existing datasets \& benchmarks.}
On the supply side two threads matter. The first is generation, a
spectrum trading fidelity against cost: DNS is the reference
standard~\cite{pope2000turbulent}, large-eddy simulation models the
unresolved scales~\cite{smagorinsky1963,germano1991,meneveau2000}, and
learned generators reproduce the statistics of the equations rather than
satisfy them~\cite{kim2020inflow,du2024confild}; TIDE sits at the DNS
end. The second is the datasets and benchmarks built on these
generators: general-purpose PDE benchmarks with predominantly 1D/2D
fluid
entries~\cite{takamoto2022pdebench,ohana2024thewell,koehler2024apebench,ren2023superbench,luo2023cfdbench},
JHTDB and the public channel-flow databases with landmark DNS
documented as one dense realization per
configuration~\cite{li2008jhtdb,lee2015channel,hoyas2006scaling}, the
classic Taylor--Green decay case, canonical precisely because its fixed
initial condition makes every run the same deterministic
trajectory~\cite{brachet1983small}, and BLASTNet~2.0 with a 3D ML-ready
corpus for compressible reacting flows, a different governing
system~\cite{chung2023blastnet}. Across these collections the held-out
variable is usually a sampling property rather than the governing
setting.

\paragraph{The position of TIDE} Table~\ref{tab:related} places TIDE in the broader landscape of turbulence datasets and PDE benchmarks. High-fidelity 3D turbulence databases provide accurate simulations, but they typically lack independent ensembles and controlled physical shifts. Machine-learning-oriented PDE benchmarks offer standardized evaluation, but they are largely limited to 1D/2D settings or different governing systems. TIDE brings these strengths together. To our knowledge, it is the first corpus of 3D \emph{incompressible} Navier--Stokes (NS) turbulence that is both DNS-verified and physically diverse. It provides independent ensembles for every configuration, a forced/decay contrast within one governing system, a certified pressure channel, and equation-level verification of the released fields. Its companion benchmark labels each transfer by the physical axis varied. It also reports physical fidelity alongside pointwise error.

\section{The TIDE Dataset}
\label{sec:dataset}

\subsection{Organization and contents}
\label{sec:overview}
\begin{figure}[t]
  \centering
  \includegraphics[width=0.92\linewidth]{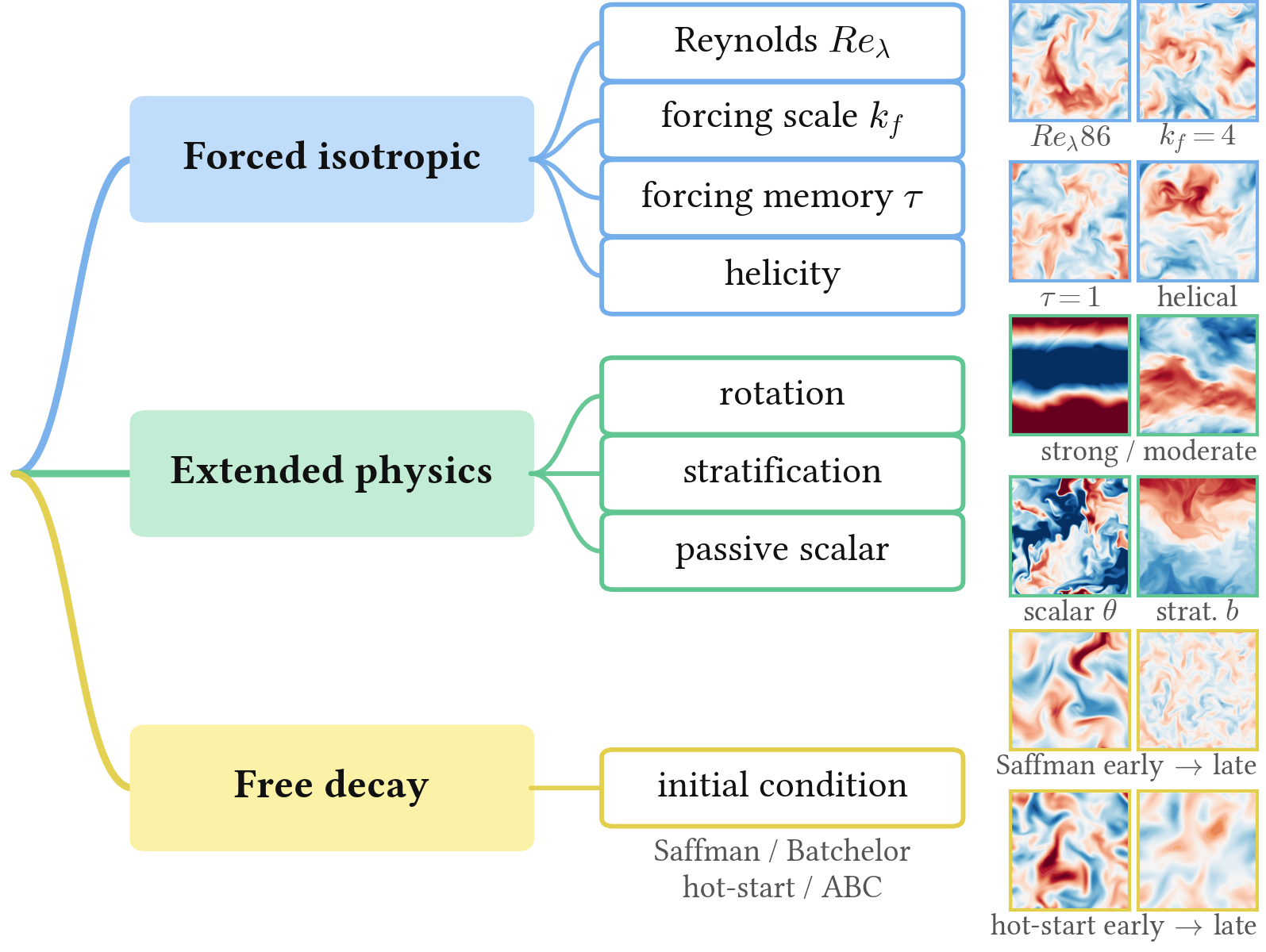}
  \caption{The taxonomy of TIDE: three regime families, defined by what
  is done to the dynamics, each varying its own physics axes;
  representative snapshots per family (decay: an early and a late
  instant). Parameters and counts in Table~\ref{tab:configs}.}
  \label{fig:taxonomy}
  \vspace{-0.5cm}
\end{figure}

Everything in TIDE is one incompressible NS system: one solver, one
grid, one acceptance standard, and only the physics varies. The top
level of Figure~\ref{fig:taxonomy} classifies what is done to the
dynamics; each family then owns its knobs, turned one at a time.
\begin{itemize}[leftmargin=1.2em,itemsep=1pt,topsep=2pt]
\item \textbf{Forced isotropic} (four axes): the equations are left
untouched and the flow is held statistically steady by stochastic
injection. The knobs are the fluid and the drive: how turbulent the
flow is (Reynolds number), where energy enters (forcing scale), how
long the drive remembers itself (forcing memory), and whether it
injects handedness (helicity).
\item \textbf{Extended physics} (three axes): exactly one ingredient is
added, a term in the momentum balance or a transported field. The knob
is what is added: rotation at two strengths, stable stratification, or
a passive scalar.
\item \textbf{Free decay} (one axis): the drive is removed. The only
knob left is what the flow starts from, four controlled initial states
that the undriven flow never forgets.
\end{itemize}
A \emph{configuration} is one setting of one knob, shipped with 8 to 16
independent realizations at $256^3$ in fp64, each passing the
acceptance standard of Section~\ref{sec:acceptance}.
Table~\ref{tab:configs} lists the configurations, measured parameters,
and counts. A physics axis denotes the physical factor being varied and
may contain multiple configurations. The eight axes yield 15
configurations: the Reynolds, forcing-scale, rotation, and
initial-condition axes contain three, two, two, and four settings,
respectively. The remaining four axes each contribute one non-default
setting, whose default level is represented by the shared
forced-isotropic reference configuration.

\begin{table}[t]
\centering
\caption{The 15 configurations: measured $Re_\lambda$ and $k_{\max}\eta$,
seeds (= independent trajectories), released frames (total across seeds),
and channels. Spectral integrals for $Re_\lambda$ use the trapezoid rule
over wavenumber shells, which reads 15--28\% lower than a sum over
shells; dissipation, $\eta$, and $k_{\max}\eta$ do not depend on it.}
\label{tab:configs}
\footnotesize
\setlength{\tabcolsep}{2.6pt}
\begin{tabular}{@{}llrrrrl@{}}
\toprule
\textbf{Config} & \textbf{Axis} & $Re_\lambda$ & $k_{\max}\eta$ & \textbf{seeds} & \textbf{frames} & \textbf{Ch.} \\
\midrule
\multicolumn{7}{@{}l}{\textit{Forced isotropic (7 configs)}} \\
$Re_\lambda 86$ (flagship) & Reynolds, $k_f{=}2$ & 84 & 1.61 & 12 & 2427 & 4 \\
$Re_\lambda 70$ & Reynolds, $k_f{=}2$ & 70 & 2.13 & 16 & 2416 & 4 \\
$Re_\lambda 55$ & Reynolds, $k_f{=}2$ & 55 & 3.00 &  8 & 1200 & 4 \\
helical          & helicity        & 83 & 1.76 &  8 & 1200 & 4 \\
$\tau{=}1$       & forcing memory  & 77 & 1.84 & 10 & 1500 & 4 \\
$k_f{=}4$        & forcing scale   & 61 & 1.69 &  8 & 1200 & 4 \\
$k_f{=}3$        & forcing scale   & 75 & 1.63 &  8 & 1200 & 4 \\
\midrule
\multicolumn{7}{@{}l}{\textit{Extended physics (4 configs)}} \\
rotating (strong)   & rotation, $\Omega_z{=}2.5$  & 285 & 2.44 & 8 & 1200 & 4 \\
rotating (moderate) & rotation, $\Omega_z{=}0.81$ & 123 & 1.81 & 8 & 1200 & 4 \\
passive scalar      & scalar, $Sc{=}1$            &  85 & 1.64 & 8 & 1089 & 5 ($\theta$) \\
stratified          & stratification, $Re_b{\approx}41$ &  86 & 1.85 & 8 & 1200 & 5 ($b$) \\
\midrule
\multicolumn{7}{@{}l}{\textit{Free decay (4 configs)}} \\
decay (hot-start)  & initial condition, $86{\to}24$ & --- & --- & 8 & 400 & 4 \\
decay (Saffman)    & initial condition, $p{=}2$ & --- & --- & 8 & 400 & 4 \\
decay (Batchelor)  & initial condition, $p{=}4$ & --- & --- & 8 & 400 & 4 \\
decay (ABC)        & initial condition, $230{\to}13$ & --- & --- & 8 & 400 & 4 \\
\midrule
\textbf{Total} & 8 axes & & & \textbf{134} & \multicolumn{2}{l}{$\approx$\textbf{2.6\,TB}} \\
\bottomrule
\end{tabular}
\\[2pt]
{\footnotesize Configuration names are nominal design labels; the tabulated
values are measured on the released frames, which is why some entries,
such as the stratified buoyancy Reynolds number and the passive-scalar
$k_{\max}\eta$, differ slightly from the per-configuration acceptance
report of Appendix~\ref{app:acceptance}, measured on each
configuration's acceptance record. Decay configurations export 50
frames per seed (window-limited by the
resolved-decay span; Table~\ref{tab:decay}); their $Re_\lambda$/$k_{\max}\eta$ are time-varying.
Store identifiers in the release map to these names via the dataset manifest.}
  \vspace{-0.5cm}
\end{table}

\subsection{Generation: solver and forcing}
\label{sec:generation}
All fields come from one pseudo-spectral solver on a $2\pi$ periodic box
at $256^3$, with $2/3$ dealiasing, an RK3 integrator with
integrating-factor viscosity, and incompressibility enforced by exact
spectral projection. Computation is fp64 throughout because fp32 leads
to a slow spectral instability at the smallest resolved scales
(Appendix~\ref{app:format}).
Statistically steady configurations are driven by Eswaran--Pope
stochastic Ornstein--Uhlenbeck (OU) forcing restricted to the largest
scales~\cite{eswaran1988forcing}. The choice follows a negative
result: deterministic band forcing is \emph{metastable} on
this grid, able to remain stationary for over a hundred eddy-turnover
times before destabilizing, so a short validation window produces a
false positive; random-phase forcing removes the metastable attractor
(Appendix~\ref{app:insights}). Every seed carries both an independent initial condition and an
independent forcing sequence, so ensemble members are fully independent
realizations; a shared forcing sequence would leave a common directional
bias that pooling over seeds cannot remove.

Two remarks fix how the axes should be read. First, the axes name a
physical property rather than an orthogonal coordinate: most settings
change one parameter relative to the reference configuration, but the
helical setting and parts of the Reynolds axis change more
(Appendix~\ref{app:discussion}). Second, initial
conditions play a double role: forced configurations forget
theirs after spin-up, so seeds provide exchangeable ensemble members,
while the four decay initial states (Saffman $p{=}2$ and Batchelor
$p{=}4$ spectra, a hot start branched from the forced $Re_\lambda 86$
trajectories, and a
maximal-helicity ABC state) decay by visibly different amounts over
the released window (Table~\ref{tab:decay}); because decay trajectories
are non-stationary, their frame windows are
truncated dynamically by the per-frame resolution gate and an energy
floor, and ensemble statistics are evaluated per decay instant rather
than time-averaged. Per-category slice strips with multiple frames are
in Appendix~\ref{app:gallery}.

\subsection{Frames, format, and curation}
\label{sec:frames}
Each seed is exported at a fixed cadence of $0.05\,T_L$, where $T_L$ is
the large-scale eddy-turnover time, typically for
150 frames and up to $250$ on the longest records, so a trajectory spans
roughly $7.5$ to $12\,T_L$. The cadence is a benchmark
design choice: at roughly one Kolmogorov time
per frame, consecutive frames differ by about $17\%$ in relative error,
enough to make forecasting
non-trivial while the fine scales remain correlated; at a coarser
cadence the fine scales decorrelate and the forecasting task degenerates
toward persistence. Configurations whose physics ends sooner, such as
free decay, carry shorter windows.

Each frame ships the velocity components and pressure, plus the scalar or
buoyancy field where the physics carries one. Fields are computed in
fp64 and stored in fp32, a precision verified against the
incompressibility gate (Appendix~\ref{app:format}). Normalization uses
frozen constants fit on each
configuration's training trajectories, with one shared velocity scale
rather than per-component scales, so the
normalization leaves component anisotropy unchanged
(Appendix~\ref{app:format}).

Frames enter the corpus through three independent gates: a per-frame
resolution gate on the stored $k_{\max}\eta$, a per-trajectory gate on
energy drift, and the per-configuration acceptance standard of
Section~\ref{sec:acceptance}. One consequence follows: the per-frame
gate preferentially drops dissipation-peak instants. It removes frames in
four of the 15 configurations, $19.1\%$ of them on $Re_\lambda 86$, where
the removed instants carry at least $1.28$ times the retained mean
dissipation, so the corpus under-samples the most intermittent events
and high-order intermittency statistics should be read as lower bounds. The corpus is released with
full documentation, a versioned DOI, and the generation, acceptance,
and benchmark code; formats, hosting, licenses, and maintenance are
detailed in Appendix~\ref{app:format}.

\section{DNS Acceptance and Verification}
\label{sec:acceptance}

Every released configuration passes a fixed, pre-specified acceptance
standard before entering the corpus: a resolution \emph{classification},
\emph{statistical} gates with accompanying literature-band checks, and
\emph{equation-level} residual checks.
Table~\ref{tab:gates} (Appendix~\ref{app:acceptance}) lists every gate
with its threshold and the measured envelope across the 15
configurations, with per-configuration values and provenance.
The acceptance scripts are themselves validated on known-answer
synthetic fields, including deliberately violating fields that must be
rejected, and the flagship configuration is reproduced independently
across hardware and software stacks with consistent statistics.

The classification decides what a configuration may claim: Class~I
fields resolve the dissipation range and license all statistics;
Class~II fields support spectra and low-order statistics only. All 15 released configurations are Class~I,
and two retained boundary anchors document where the $256^3$ ceiling
lies (Appendix~\ref{app:discussion}). Each statistical gate is judged as
pass or fail and failing one rejects the configuration, with any
admitted exception disclosed per configuration in
Appendix~\ref{app:acceptance}; together they protect
resolution of the smallest scales with a clean spectral tail
(Figure~\ref{fig:spectra}, Appendix~\ref{app:acceptance}), separation of the energy-containing scales
from the box, a steady record whose injection balances dissipation, and
the isotropy, incompressibility, and derivative skewness of a physical
cascade. For rotation, stratification, and free decay, anisotropy or
non-stationarity is the physics under study, so the isotropy and
stationarity gates do not apply, while resolution, incompressibility,
and equation-level gates remain hard; pooling conventions and the
extension-specific gates are in Appendix~\ref{app:acceptance}.

\looseness=-1
Statistical gates cannot exclude compensating errors, so four
equation-level checks verify the discrete governing equations directly
(Appendix~\ref{app:dgroup}): the \emph{divergence} residual is at solver
precision under exact projection; the \emph{momentum} residual, on frame
triplets with frozen forcing, contains only time truncation, which a
\emph{step-halving} check confirms at second order; and the
\emph{pressure} check verifies the shipped pressure channel against its
Poisson equation.

\section{Benchmark Design}
\label{sec:benchmark}

The benchmark is the measuring instrument for the corpus structure of
Section~\ref{sec:dataset}: five reference tasks, one training and
evaluation protocol, a deterministic evaluation slice, generalization
splits along the physics axes, and a three-axis metric suite.

\subsection{Tasks and experimental setup}
Every task is the same object: a single-step map
$f:(C_{\mathrm{in}},256^3)\rightarrow(C_{\mathrm{out}},256^3)$ learned by
regression, the five differing only in how the input--target pair is
built from the released frames (Table~\ref{tab:tasks}). Only P1 advances
time; P2--P5 map between fields at the same instant. Not every baseline runs
on every task: forecasting carries four learned models, each single-frame
task three or four, and DeepONet-3D runs on pressure recovery only
(Appendix~\ref{app:results}). Two
further tasks, denoising and physics-parameter inversion, are specified
in Appendix~\ref{app:planned-tasks} and are not part of the reference
results.

\subsubsection{P1: Forecasting.}
The pair is $(u(t),\,u(t{+}\Delta t))$ at the corpus cadence
$\Delta t=0.05\,T_L$, roughly one Kolmogorov time, over all stored
channels. Training uses only this single step, sampled by a sliding
window whose start advances 4 frames (frame stride 4), giving about 110
pairs on most forced configurations; consecutive frames are highly
correlated, and denser sampling is studied in
Appendix~\ref{app:recipe}. Models predict the
\emph{increment}, $\hat u(t{+}\Delta t)=u(t)+f(u(t))$: consecutive
frames are far more similar than either is to the mean, so direct
regression of the absolute field admits a near-constant minimizer close
to the sample mean, whereas under residual
prediction the persistence solution is the zero output and the model
learns the correction to it. Evaluation iterates the trained single-step
map autoregressively for 20 steps ($1.0\,T_L$), a horizon never seen in
training, so rollout is \emph{zero-shot}. Evaluation windows start every 8
frames, up to 51 per configuration, fewer on the shorter scalar and
decay records. On the flagship, whose seven test trajectories are
longer, the 51 windows cover about two of them; scoring all 135 changes
the models by $-0.6\%$ to $+3.9\%$ (Appendix~\ref{app:extended-results}). Persistence, copying the input forward, is the trivial reference; the
equations-informed integrator, a classical spectral solver given the
true equations, is the informed one.

\subsubsection{P2: Super-resolution.}
The input is the true frame downsampled $4\times$ by sharp spectral
truncation, a clean anti-aliased coarsening rather than pixel averaging,
and interpolated back onto the $256^3$ grid; the target is the original
frame.
Sampling takes 20 equally spaced frames per trajectory. The trivial
reference is spectral interpolation alone, i.e.\ the input itself.

\subsubsection{P3: Sparse reconstruction.}
A random 5\% of grid points is kept and the rest set to zero; the target
is the full field. The observation mask is fixed per sample, so every
model is scored against the identical observation set.
Sampling matches P2 (20 frames per trajectory); the trivial reference is
identity.

\subsubsection{P4: Pressure recovery.}
The input is the three velocity components and the target is the
pressure field of the same frame; the stored pressure channel is
deliberately withheld from the input. Because pressure is determined
instantaneously by an elliptic equation, the value at one point depends
on the entire field, so the task probes whether a model can learn a
global, non-local operator. Its reference is \emph{informed} rather than trivial: the exact spectral
Poisson solve, the same operator that
generated the corpus pressure channel, reproduces it to nRMSE
$\approx\!1.1\times10^{-6}$, and a model's distance from it measures how
much of the operator it has not learned. Evaluation adds the Poisson
residual, substituting the predicted pressure back into its defining
equation.

\subsubsection{P5: Subgrid-stress closure.}
The input is the Gaussian-filtered frame and the target is the six
independent components of the subgrid stress
$\tau_{ij}=\overline{u_iu_j}-\bar u_i\bar u_j$, computed from the
unfiltered velocity, the classic a-priori test of large-eddy simulation.
The trivial reference is dynamic Smagorinsky with the coefficient set by
the Germano identity. Beyond pointwise error, the task scores the
stress correlation and the backscatter fraction: an eddy-viscosity
closure is dissipative by construction and returns exactly zero
backscatter, while real turbulence transfers energy upscale part of the
time, so reproducing a non-zero backscatter fraction is something the
classical form cannot do at any coefficient.

\subsubsection{Training and evaluation protocol.}
\label{sec:native-eval}
Evaluation operates on the full $256^3$ field with no tiling, as does
training for every model except the Spectral U-Net, whose full-field
backward pass exceeds memory and which trains on $128^3$ crops while
still being evaluated natively (Appendix~\ref{app:baselines}). Evaluating natively is a well-posedness
requirement: the physical metrics rest on a periodic FFT, and a
sub-block of a periodic box is not itself periodic; a tiled alternative,
evaluated as an ablation, injects seam errors that autoregression
amplifies (Appendix~\ref{app:results-ablation}).

Models see a single frame, which the deterministic part of the dynamics
justifies: pressure is fixed instantaneously by an elliptic equation, so
the velocity field carries no explicit memory. The stochastic drive is
the exception, since its Ornstein--Uhlenbeck state is not an input and
cannot be read off one frame; within a configuration the model learns
that drive only in expectation, and across the forced/decay boundary the
omission becomes decisive (G4 below). Single-frame input is therefore a
protocol choice rather than a consequence of the equations; a short
history and an explicit forcing channel, which partially identify the
drive, are evaluated in Appendix~\ref{app:conditioning}. One recipe is fixed
across all baselines and tasks, AdamW at learning rate $10^{-3}$ with
cosine annealing over 25 epochs (full settings in
Appendix~\ref{app:protocol}), since per-model tuning would fold tuning
budget into the comparison; a variant with the learning rate selected
on validation rollouts is reported alongside (Appendix~\ref{app:recipe}). Inputs are scaled by the
frozen normalization constants of Section~\ref{sec:frames}.

Reference results are computed on a deterministic \emph{benchmark
slice}: per configuration, three training, one validation, and three
test trajectories (seven on the flagship), ranked by a model-independent quality score with published weights and
splits disjoint by construction (Appendix~\ref{app:slice}). Sampling
uncertainty should be read against an effective sample size of
$N_{\mathrm{eff}}\approx9$, set by trajectory count and length rather
than by sampling density (Appendix~\ref{app:slice}); reference results
are medians over five training seeds, three for Transolver and the
Spectral U-Net on $\tau{=}1$ (Section~\ref{sec:rollout}).

\subsubsection{Generalization across physical regimes.}
The tasks above are trained and evaluated within a configuration. The
controlled axes of Section~\ref{sec:dataset} support a second mode of
evaluation: train the forecasting task (P1) on one set of
configurations, hold a target regime out entirely, and measure zero-shot
transfer onto it. Table~\ref{tab:ggroup} defines five such transfers,
G1--G5, each named by the physical axis it varies. \textbf{G1},
cross-$Re$, varies viscosity alone between $Re_\lambda 70$ and $86$,
while $Re_\lambda 55$ also differs in forcing variance; G1 and G3 are
scored separately for each source, and G1 also with a jointly trained
model. \textbf{G2}, \textbf{G3},
and \textbf{G5} move several physical factors at once and are reported as
regime transfers without single-factor attribution, while \textbf{G4} is
not scored as a transfer at all, for the reason below. For G2, G3, and
G5 the property defining the target regime is present in the input
field, so the operators can in principle adapt and the obstacle is
training coverage. G5 is bounded by channels: a four-channel model
cannot be applied zero-shot to the five-channel scalar and stratified
corpora, so it runs on the rotating configurations.

\textbf{G4}, the forced$\rightarrow$decay axis, is different in kind.
Removing the forcing changes no term of
the Navier--Stokes equations, yet the drive is exogenous: the
instantaneous forcing cannot be read off one frame, and a snapshot early
in a decay is statistically indistinguishable from a forced one, so a
forced-trained operator continues to advance it in the
injection regime whether or not it has internalized the dynamics. The
failure is evidence of a missing conditioning input, not of how well the
physics was learned; to our knowledge no 3D neural operator takes the
energy injection as an explicit input (Appendix~\ref{app:conditioning}
evaluates such a channel), and forcing transfer has been
demonstrated only in 2D with a learned-correction
hybrid~\cite{kochkov2021ml}. For the same reason the axis is not scored
as a transfer error, which would be governed by the reference
trajectories rather than by the operator; G4 is evaluated within each
flow, against the non-learned references
(Section~\ref{sec:results}).

\subsection{Baselines and metrics}
Five learned baselines span the dominant operator families:
\textbf{FNO3d}, a Tucker-factorized FNO (\textbf{TFNO}), a
\textbf{Spectral U-Net}, \textbf{Transolver}, and
\textbf{DeepONet-3D} (Table~\ref{tab:baselines}). TFNO and the Spectral
U-Net are audited variants of published designs, with every deviation
tabulated in Appendix~\ref{app:deviations}. Comparability is defined by
matched capacity: the four spectral and attention models sit within
$28$--$30.3$\,M parameters, matched upward by enlarging the smaller
models, with DeepONet-3D at the memory ceiling of its dense trunk. TFNO
retains $18^3$ Fourier modes against $12^3$ for FNO3d and the Spectral
U-Net.
A median run takes under an hour on one consumer GPU and a forecasting
run $1.25$--$2.25$\,h, so the full reference matrix is reproducible on a
single workstation
(Appendix~\ref{app:compute}). Two kinds of non-learning reference frame every task:
\emph{trivial references} (persistence, identity, spectral
interpolation, dynamic Smagorinsky), which a useful model must beat, and
\emph{informed references} (the exact Poisson solve, the
equations-informed integrator), which mark what known physics attains.

The leaderboard reports three axes, because a single averaged error is
blind to the failure mode that dominates rollout: pointwise nRMSE
(normalized root-mean-square error)
averaged over the rollout, the
final-step enstrophy ratio (target $1.0$), and final-step high-band
spectral error.
A per-case skill score,
$1-\mathrm{nRMSE}/\allowbreak\mathrm{nRMSE}_{\mathrm{persistence}}$, places each
model against its own flow's persistence, since absolute errors
are not comparable across configurations, and the effective prediction
time reports where a rollout crosses nRMSE $0.3$. Subgrid stress adds
the stress correlation and the backscatter fraction, for the structural
reason given with P5; full definitions are in
Appendix~\ref{app:metrics}.

\section{Results}
\label{sec:results}

\begin{table*}[t]
\centering
\caption{Forecasting (P1) per configuration, median over training seeds:
nRMSE averaged over the 20 rollout steps (lower is better), followed by
the number of diverging seeds $(k/n)$ when nonzero / final-step
enstrophy ratio (target $1.0$) / effective prediction time (EPT, frames
until nRMSE crosses $0.3$; one frame $=0.05\,T_L$, higher is better).
Five seeds per cell (three for Transolver and the Spectral U-Net on
$\tau{=}1$), trained under the reference recipe of
Section~\ref{sec:benchmark} (frame stride 4); zero-shot 20-step rollout at native
$256^3$ on 51 windows per configuration; seed ranges in
Table~\ref{tab:app-seeds}. A rollout diverges when its 20-step mean
nRMSE exceeds 1. The persistence and equations columns (nRMSE) bracket
the task: persistence is the error a useful model must fall below, the
equations column$^{\P}$ the error no benchmarked model reaches.
\underline{Underlined}: lowest median nRMSE; $^{\star}$: separable from
the second-lowest model on every seed the two share (paired trajectory-level 95\% bootstrap CI excludes 0). $^{\P}$The
equations-informed integrator, fp32, no access to the forcing
realization, eight windows.}
\label{tab:main-results}
\small
\setlength{\tabcolsep}{4pt}
\begin{tabular}{@{}lcccccc@{}}
\toprule
\textbf{Config} & \textbf{FNO3d} & \textbf{TFNO} & \textbf{Transolver} & \textbf{Spectral U-Net} & \textbf{persistence} & \textbf{equations} \\
\midrule
$k_f{=}3$ & \underline{0.779} / 3.4 / 2.0 & 14.0 (5/5) / $5{\times}10^{4}$ / 2.1 & 0.810 / 8.8 / 0.5 & 0.830 / 17.2 / 1.0 & 0.841 & 0.410 \\
$k_f{=}4$ & \underline{0.672}$^{\star}$ / 1.4 / 2.2 & 0.763 / 0.5 / 0.8 & 0.799 / 7.0 / 0.8 & 0.851 / 17.8 / 1.2 & 0.822 & 0.403 \\
$\tau{=}1$ & 1.149 (4/5) / 12.9 / 1.5 & 1.838 (5/5) / 279.7 / 1.9 & \underline{0.843} / 19.9 / 0.2 & 0.860 / 36.6 / 0.7 & 0.890 & 0.488 \\
$Re_\lambda 55$ & 0.905 (1/5) / 12.3 / 2.8 & 0.786 (1/5) / 48.4 / 1.0 & 0.803 / 14.6 / 0.8 & \underline{0.767}$^{\star}$ / 16.7 / 1.3 & 0.831 & 0.411 \\
$Re_\lambda 70$ & 0.929 (2/5) / 12.8 / 2.6 & 9.97 (4/5) / $5{\times}10^{4}$ / 2.0 & \underline{0.789} / 16.1 / 0.6 & 0.804 / 38.7 / 1.0 & 0.822 & 0.405 \\
$Re_\lambda 86$ & 1.065 (4/5) / 12.4 / 2.2 & 1.371 (3/5) / 104.9 / 3.0 & \underline{0.819} / 63.8 / 0.4 & 0.820 / 17.7 / 0.9 & 0.853 & 0.441 \\
\bottomrule
\end{tabular}
\vspace{-0.25cm}
\end{table*}

\looseness=-1
Four findings organize this section: under the reference recipe,
learned operators beat persistence only modestly and stay about twice
the error of the equations-informed integrator; low pointwise error does
not imply physical fidelity; rollout stability depends on configuration,
training seed, and training-data density rather than on architecture
alone; and the forced-to-decay axis isolates what an operator is
conditioned on. Table~\ref{tab:main-results} reports the forecasting
leaderboard per configuration as medians over five training seeds
(three for Transolver and the Spectral U-Net on $\tau{=}1$).
Matrices for the remaining tasks and configurations are in
Appendix~\ref{app:results}. Evaluation is zero-shot under the protocol
of Section~\ref{sec:benchmark}. Two models are separable on a seed when
the paired trajectory-level 95\% bootstrap interval of their difference
excludes zero, and an ordering is called separable when this holds on
every seed the two models share.

\subsection{Accuracy across operator families}
\looseness=-1
No one model leads on every task, and the per-task winners differ
(Appendix~\ref{app:results}). On forecasting, Transolver is the only
model whose median skill score is positive on all six configurations
($+0.027$ to $+0.053$), while FNO3d records the highest single
configuration ($k_f{=}4$, $+0.182$, separable from the runner-up on
every seed) yet diverges on four of five seeds on $\tau{=}1$ (median
skill $-0.291$). On super-resolution FNO3d improves on the spectral-interpolation
reference on every main-table configuration, by $0.012$--$0.048$ in
nRMSE; on the eight specialized regimes the outcome is parity, six
sitting within $0.015$ of the reference in either direction, and the
reference clearly ahead on the other two (strong rotation, quasi-steady
ABC). The learned-model ordering is nonetheless
the suite's most stable, FNO3d leading all fourteen configurations that carry the task, unlike
forecasting (Appendix~\ref{app:results}).

Two tasks are more informative than their numbers alone. On pressure
recovery every network sits well above the exact Poisson reference, with
DeepONet-3D at nRMSE $\approx 1.0$, the level of a zero field: its global
branch encoding discards spatial arrangement, so the minimizer of that
parameterization is near the sample mean, a property of the encoding
rather than a training failure. The pattern inverts on subgrid-stress
closure,
where no exact operator exists: FNO3d and TFNO reach stress
correlations of $0.45$--$0.54$ against dynamic Smagorinsky's
$0.16$--$0.17$ and report a non-zero backscatter fraction, which an
eddy-viscosity closure returns as exactly zero by construction. The
Spectral U-Net does not share that gain, reaching stress correlations of
$0.15$--$0.28$ and falling below dynamic Smagorinsky on the lowest
Reynolds configuration, and on the seven extended configurations where both run TFNO stays
below nRMSE 1 ($0.61$--$0.84$) while the Spectral U-Net stays above it
($1.26$--$11.0$), so architectures separate within one task and not
only across tasks. Both failures trace to what an architecture is allowed to see rather
than to how hard the task is, and both are invisible from forecasting
alone, which is what a multi-task suite buys: scored on any single task
the benchmark would report a different leader.

\subsection{Physical fidelity vs.\ pointwise error}
\label{sec:three-axis}
\begin{figure}[t]
  \centering
  \includegraphics[width=0.85\linewidth]{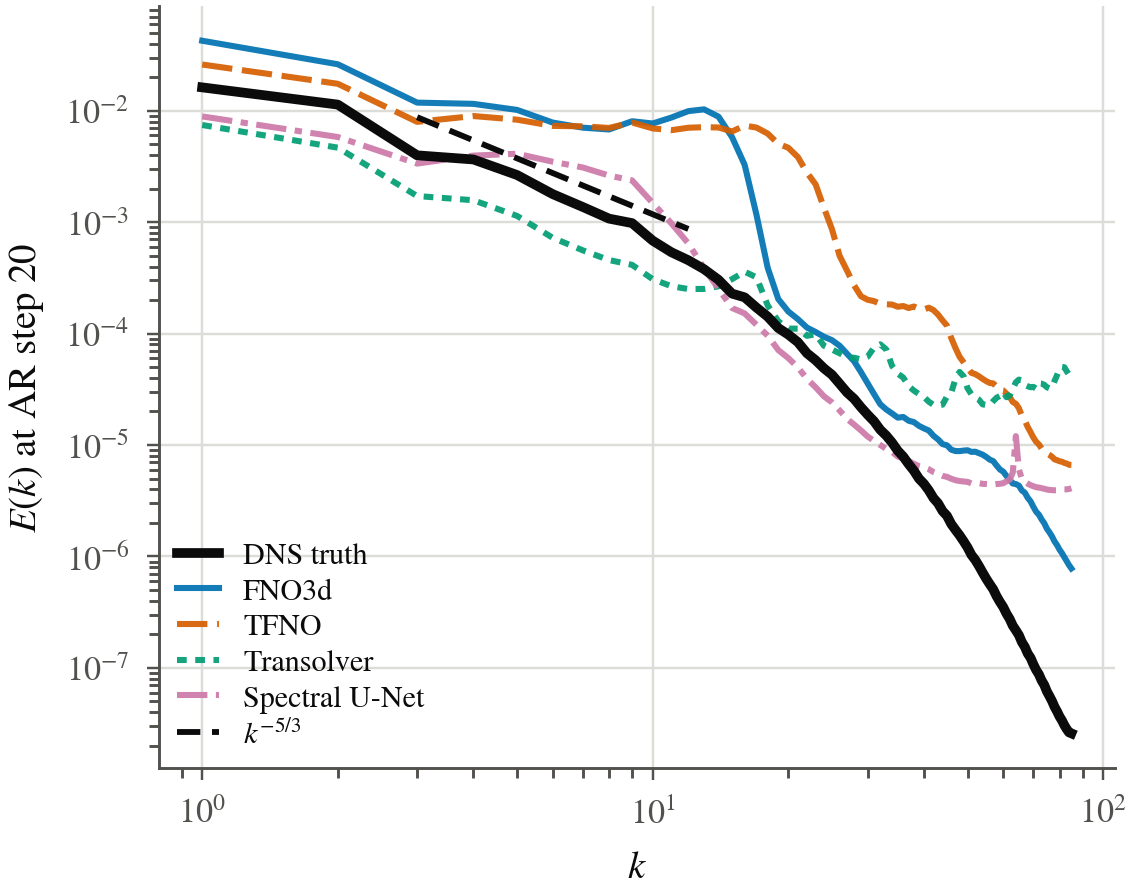}
   \vspace{-0.3cm}
  \caption{Predicted vs.\ true energy spectra at rollout step 20
  ($Re_\lambda 86$, seed 0, mean over test samples).}

  \label{fig:rollout-spectra}
\end{figure}

The three axes of Table~\ref{tab:main-results} are led by different
models, and we report the disagreement rather than a ranking. On
$k_f{=}3$ Transolver's median nRMSE is $0.031$ above FNO3d's while
its median high-band spectral error is fifty times FNO3d's ($442$ vs.\
$8.8$), and averaged over the six configurations the gap is
thirty-five-fold ($607$ vs.\ $17$; per-configuration values in
Table~\ref{tab:app-highband}); on the flagship $Re_\lambda 86$
configuration Transolver has the lowest median nRMSE ($0.819$) while
its median final-step enstrophy ratio is $64$, well over an order of
magnitude above the target of $1.0$. Figure~\ref{fig:rollout-spectra}
shows the corresponding spectra: at rollout step 20 every model departs
from the DNS spectrum well before the resolution scale, flattening into
a high-wavenumber floor one to three orders of magnitude above truth,
while Transolver additionally loses energy at the largest scales. One pair makes the point directly: on $Re_\lambda 86$ Transolver and the
Spectral U-Net differ by $0.001$ in median nRMSE, while their median
enstrophy ratios differ by a factor of $3.6$ ($64$ vs.\ $18$). The distortions also land in different bands, one model spiking only at
the highest wavenumbers while another spreads a broadband excess
(Appendix~\ref{app:extended-results}). Persistence makes the converse point, preserving small-scale
statistics by construction (enstrophy ratio $0.96$--$1.00$, near-zero
high-band error) with zero skill by definition. The disagreement is the
common case: in 13 of the 15 cells of Table~\ref{tab:main-results}
whose median nRMSE is below persistence, the median enstrophy ratio
departs from the target by more than a factor of two. Pointwise error, small-scale physical health, and the location of the
spectral distortion are therefore not substitutes: a single averaged
error is free to rank a degraded field above a healthy one.

\subsection{Rollout stability}
\label{sec:rollout}
Single-step accuracy does not predict step-20 behavior. On
$Re_\lambda 86$ TFNO has the lowest first-step error of the four models
($0.45$ of persistence's at seed 0) yet diverges on three of five seeds,
while Transolver starts closest to persistence ($0.96$) and never
diverges. Divergence arises in the accumulation over the horizon rather
than in the first step (Figure~\ref{fig:rollout-curves}). Across the
six configurations FNO3d diverges on $0/5$ to $4/5$ seeds and TFNO on
$0/5$ to $5/5$ (Table~\ref{tab:main-results}; ranges in
Table~\ref{tab:app-seeds}), while Transolver and the Spectral U-Net
never diverge on the forced configurations.

\looseness=-1
The counts follow where a model sits relative to persistence. Every run
starts on a persistence plateau, the solution whose increment is near
zero, and the reference budget does not move every model off it: on the
forced configurations Transolver's first-step error stays at
$0.93$--$1.00$ of persistence's, and TFNO on $k_f{=}4$ at
$0.993$--$0.996$ on all five seeds. These $0/5$ counts record models
that have barely left persistence, which is a different finding from a
stable operator. Training on every frame of the same trajectories
(frame stride 1, four times the training pairs per epoch; seed 0,
Table~\ref{tab:app-stride1}) moves the cells off the plateau: the
first-step error relative to persistence falls in all 14 cells where it
was measured, and the flagship Transolver improves from $0.823$ to
$0.713$ in 20-step error on full trajectories (lower on 7/7). Denser
sampling does not remove divergence: four stride-1 cells end above 1,
among them TFNO on $k_f{=}4$ ($1.59$), which sits on the plateau on all
five stride-4 seeds. Whether a cell leaves the plateau, and whether it
then diverges, therefore depends on how densely training samples the
trajectories as much as on the seed, and Table~\ref{tab:main-results}
measures the reference recipe rather than the architectures in
isolation.

Re-realization sets the resolution of the comparison: on four of the
six configurations the two best models are not separable on every seed
(Table~\ref{tab:main-results}), and on the
flagship they change order under full-trajectory scoring
(Appendix~\ref{app:extended-results}). Seed spreads are smallest for
Transolver (at most $0.005$ on every forced configuration) and reach
$0.032$ for the Spectral U-Net. Regime diversity extends the same
caution: Transolver, the most seed-stable model under forcing, diverges
on the hot-start decay on all three seeds under the decay divergence
rule, and the Spectral U-Net diverges on all four free-decay families
with the reference seed (Table~\ref{tab:app-rollout-ext}).

\subsection{Generalization along controlled axes}
\begin{table}[t]\centering
\caption{Zero-shot transfer onto the $Re_\lambda 86$ test set: 20-step
nRMSE on all 135 windows of the seven test trajectories, under the
target configuration's / the source model's input normalization.
Sources are seed-0 models trained under the reference recipe; TFNO's
$Re_\lambda 70$ and $k_f{=}3$ sources already diverge in distribution.}
\label{tab:app-ood}\footnotesize\setlength{\tabcolsep}{2.6pt}
\begin{tabular}{@{}lcccc@{}}
\toprule
Source & FNO3d & TFNO & Transolver & Spectral U-Net \\
\midrule
$Re_\lambda 70$ & 0.948 / 0.999 & 8.23 / 9.20 & 0.828 / 0.828 & 0.818 / 0.802 \\
$k_f{=}3$ & 0.856 / 1.165 & 33.0 / 113 & 0.831 / 0.835 & 0.854 / 0.778 \\
$k_f{=}4$ & 0.812 / 1.222 & 0.801 / 0.802 & 0.840 / 0.845 & 0.827 / 0.763 \\
$\tau{=}1$ & 1.029 / 1.214 & 1.572 / 1.913 & 0.814 / 0.819 & 0.872 / 0.802 \\
\midrule
in-distribution & 1.022 & 0.969 & 0.823 & 0.816 \\
persistence & 0.860 & 0.860 & 0.860 & 0.860 \\
\bottomrule
\end{tabular}
\vspace{-0.3cm}
\end{table}

\looseness=-1
We report zero-shot transfer in three directions: onto the flagship
$Re_\lambda 86$ test set (Table~\ref{tab:app-ood}, full trajectories),
onto $k_f{=}4$ along the forcing axis, and onto the rotating sets in the
reverse of the specialization direction (Appendix~\ref{app:results}).
Ordering the sources by physical distance largely reproduces the
ordering of transfer error: from the nearest Reynolds neighbor
Transolver and the Spectral U-Net transfer at their in-distribution
level ($0.828$ vs.\ $0.823$ and $0.818$ vs.\ $0.816$), while strongly
specialized sources score above persistence, strong rotation reaching
$1.13$--$1.63$. Cutting the rotation rate by a factor of three moves
Transolver from $1.31$ to $0.85$ while FNO3d stays near $1.13$, the two
levels differing in kind (ensemble-pooled two-dimensional energy
fraction $0.95$ against $0.64$); on the cross-physics probe only
Transolver beats persistence, by a margin small relative to the seed
spread. FNO3d, weak in distribution on the flagship, scores lower when
transferred from the forcing-scale neighbors ($0.81$--$0.86$ against
$1.02$), but only under the target configuration's input normalization:
under the source model's own normalization the same entries rise to
$1.16$--$1.22$, so the normalization is part of the transfer protocol
and both are tabulated.

The reverse of the rotation transfer does not mirror it. An operator
trained on a generic forced flow can move onto strong rotation
essentially without loss: Transolver reaches $0.485$--$0.501$ against
its in-distribution $0.488$, and under the source model's normalization
Transolver and the Spectral U-Net from $k_f{=}3$ score $0.323$ and
$0.430$, below their in-distribution errors. The rotation-specialized
operator moved onto the generic flow reaches $1.31$: physical
specialization transfers in one direction only.

\looseness=-1
Joint training tests whether the scored gaps are a matter of training
coverage (Appendix~\ref{app:joint}). On none of the three axes tested
is the joint model better than both the reference single-source model
and a control matched in optimizer steps under both input
normalizations: on the Reynolds axis the Transolver joint model is not
separable from its single source and FNO3d is worse ($+0.134$). On the
rotation axis the Transolver joint model beats the step-matched control
but not the reference single-source model, and adding hot-start data to
flagship training at matched steps changes the decay error by $+0.0003$
for Transolver. No single cause accounts for the gaps on all three
axes. The forced$\rightarrow$decay axis (G4) is not
scored as a transfer, for the reasons fixed in
Section~\ref{sec:benchmark}. Its conditioning question is measured
within each flow instead. Across four forced configurations the
equations-informed integrator's 20-step error is $0.40$--$0.49$ with no forcing input,
$0.24$--$0.39$ with the current forcing state estimated from two past
frames and propagated by its Ornstein--Uhlenbeck conditional mean, and
about $0.001$ with the true forcing. The removable share is $35$--$41\%$
for the $\tau{=}2$ configurations and $20\%$ for $\tau{=}1$, the
shortest forcing memory, and the remainder is an upper bound on the part
no input can remove. Learned operators given the same information gain
less consistently: four input frames help Transolver on
$Re_\lambda 86$ and hurt FNO3d, and an explicit forcing channel lowers
error against its zeroed counterpart on every seed of the three
model--configuration pairs trained with three seeds, within the zeroed channel's cross-seed spread for FNO3d
(Appendix~\ref{app:conditioning}). On the
hot-start decay itself, where no forcing is hidden, the integrator is
near-exact (nRMSE below $10^{-4}$), while on the three non-quasi-steady
decay families the learned operators span $0.79$ to $1.56$ against
persistence values of $0.80$ to $1.02$.

\looseness=-1
Persistence holds a six-configuration mean nRMSE of $0.843$ and the
equations-informed integrator reaches $0.426$. Under the reference
recipe, no baseline improves on persistence by more than $4\%$ in
six-configuration mean skill (Transolver, $+0.038$), though cells
qualify that (FNO3d reaches $+0.182$ on $k_f{=}4$; the quasi-steady ABC
case leaves almost no headroom). The recipe is not a ceiling: a
four-step unrolled loss lowers FNO3d on $Re_\lambda 70$ from $0.929$ to
$0.703$, and denser frame sampling moves the flagship Transolver to
$0.713$ (Appendix~\ref{app:recipe}). Even so, at least half of the
demonstrated predictability, and more than four fifths on average, is claimed
by neither reference nor any operator we benchmarked. That unclaimed
interval, rather than the ordering within it, is what the benchmark
leaves open.

\section{Conclusion and Future Work} In this paper, we introduced TIDE, a physically diverse, DNS-verified corpus and benchmark for three-dimensional incompressible turbulence. It provides independent ensembles, controlled physical shifts, equation-level validation, and standardized generalization splits. Under a fixed reference recipe, current learned operators only modestly outperform persistence and remain substantially less accurate than a spectral solver supplied with the governing equations, although unrolled training and denser sampling move individual cells well below persistence. Lower pointwise error can mask distorted small-scale dynamics, and whether a rollout stays stable depends on the configuration, the seed, and the training density. Joint training across neighboring regimes does not close the transfer gaps we measure. The forced-to-decay contrast bounds what the missing forcing state contributes: an estimate of the current forcing state removes 20--41\% of a forcing-free solver's error, while history or forcing inputs help only some architectures. These findings indicate that progress in scientific machine learning requires more than predictive accuracy. Reliable models also need to preserve physical fidelity, remain stable during rollout, and receive appropriate conditioning information. TIDE is currently limited to moderate Reynolds numbers, a periodic domain, and one pseudo-spectral solver family, and Appendix~\ref{app:discussion} states each boundary in full. Future work should extend the corpus to higher Reynolds numbers and broader physical regimes and develop rollout-aware and conditioned operators. More broadly, we encourage the community to use or extend TIDE for rigorous model comparison, as well as for developing large-scale pretrained models of three-dimensional turbulence.

\clearpage

\bibliographystyle{ACM-Reference-Format}
\bibliography{references}

\appendix
\setlength{\emergencystretch}{2em}

\section{Per-Configuration DNS Acceptance Tables}
\label{app:acceptance}
Table~\ref{tab:gates} lists every acceptance gate with its threshold and
the measured envelope over the configurations it applies to
(Section~\ref{sec:acceptance}).
Table~\ref{tab:app-acceptance} summarizes the acceptance verdict and the
governing gates for all 15 released configurations, and
Table~\ref{tab:app-dgroup} lists the equation-level D-group measurements.
The item-by-item reports behind both tables, each carrying its own
provenance stamp (git hash, date, machine, library versions), ship with
the corpus release; the values here are parsed from those reports rather
than re-derived. The released reports label the statistical gates
A1--A13 and the equation-level checks D1--D4; those labels are retained
in this appendix for traceability. The literature-band quantities
(Kolmogorov constant, dissipation coefficient, derivative flatness,
4/5-law proximity) are reported rather than gated, since several bands
are defined only at higher Reynolds number; Table~\ref{tab:lit-bands}
tabulates all four for the seven forced configurations. The derivative
flatness and the 4/5-law proximity fall inside their bands on all seven.
For the Kolmogorov constant and the dissipation coefficient the standard
attaches no verdict at this Reynolds range: the compensated-spectrum
plateau sits in the bottleneck region rather than in an inertial range,
and $C_\varepsilon$ rises above its high-$Re$ band through known
finite-Reynolds
corrections~\cite{sreenivasan1998update,bos2007spectral}; the measured
values follow both trends.

\begin{table}[H]
\centering
\caption{Literature-band quantities for the seven forced configurations
(reported, not gated): Kolmogorov constant $C_K$ with its plateau
location, dissipation coefficient $C_\varepsilon$, derivative flatness
$F$, and 4/5-law proximity $\max_r[-S_3/(\varepsilon r)]$.}
\label{tab:lit-bands}
\footnotesize
\setlength{\tabcolsep}{4pt}
\resizebox{\linewidth}{!}{\begin{tabular}{@{}lcccc@{}}
\toprule
\textbf{Config} & $C_K$ (plateau $k\eta$) & $C_\varepsilon$ & $F$ & \textbf{4/5-law} \\
\midrule
literature band & 1.45--1.79$^{a}$ & n/a ($Re_\lambda{<}100$)$^{b}$ & 4--8 & 0.5--0.85 \\
\midrule
$Re_\lambda 55$ & 2.36 (0.14) & 0.665 & $4.70\pm0.23$ & $0.62\pm0.17$ \\
$Re_\lambda 70$ & 2.46 (0.13) & 0.599 & $4.87\pm0.22$ & $0.64\pm0.14$ \\
$Re_\lambda 86$ & 2.40 (0.17) & 0.572 & $5.34\pm0.31$ & $0.63\pm0.09$ \\
$k_f{=}3$       & 2.43 (0.17) & 0.552 & $4.90\pm0.18$ & $0.58\pm0.06$ \\
$k_f{=}4$       & 2.41 (0.18) & 0.602 & $4.71\pm0.14$ & $0.59\pm0.04$ \\
$\tau{=}1$      & 2.48 (0.11) & 0.565 & $5.24\pm0.26$ & $0.60\pm0.10$ \\
helical         & 2.36 (0.10) & 0.573 & $5.21\pm0.31$ & $0.66\pm0.13$ \\
\bottomrule
\end{tabular}}
\\[2pt]
{\footnotesize $^{a}$The band presumes an inertial-range plateau
($k\eta\approx0.02$--$0.05$, requiring $Re_\lambda>140$); the measured
plateaus sit at $k\eta=0.10$--$0.18$, inside the bottleneck range, so
the value is reported without a verdict~\cite{yeung1997universality}.
$^{b}$The $C_\varepsilon$ band is defined for $Re_\lambda\ge100$; below
that, low-$Re$ corrections lift $C_\varepsilon$ and the standard
prescribes comparison against the literature trend
only~\cite{sreenivasan1998update}, with the integral scale of the
scale-separation gate.}
\end{table}

High-variance signed
statistics, such as the velocity cross-correlation, are judged on
ensemble-pooled values: single realizations fluctuate by several
percent, reaching $6\%$ on the largest ensembles, while the pooled
cross-correlations of the released frame sets span $0.9$--$1.8\%$
against the $2\%$ gate and the component-energy anisotropy
$0.9$--$3.9\%$ against $5\%$. The pool estimates the flow's isotropy, and pooling is unbiased
only because every seed carries an independent forcing sequence; studies
sensitive to component anisotropy should likewise pool over
realizations. One configuration is admitted with its pooled closure
above the nominal threshold, noted in Table~\ref{tab:app-acceptance}.

The gates compressed to ``pass'' in Table~\ref{tab:gates} have measured
envelopes. Across the forced configurations the box-to-integral-scale
ratio spans $5.2$--$8.5$ ($3.0$ and $4.1$ under rotation), the advective
CFL stays at or below $0.40$ against a $0.5$ gate and the dissipative
step $\Delta t/\tau_\eta$ at $0.010$--$0.016$ against $0.05$, spin-up
discards $12$--$21$ energy-based eddy-turnover times $T_E$ before
sampling with averaging windows of
$7$--$234\,T_E$, and the derivative skewness sits at $-0.50$ to $-0.52$
on the forced records, $-0.48$ to $-0.51$ for the scalar, stratified,
and moderate-rotation cases, $-0.42$ to $-0.50$ on the free-decay
families, and $-0.23$ under strong rotation, where rotation suppresses
the cascade. The gates bind in practice: the forcing-memory axis has two points
rather than three because one candidate configuration passed admission
on its long-trajectory statistics yet failed the stationarity gate on
its frame set, and one scalar realization was truncated by the per-frame
resolution gate, shortening its released record.

\begin{table}[t]
\centering
\caption{Every hard acceptance gate: threshold and measured envelope over
the configurations the gate applies to (per-configuration values in
Tables~\ref{tab:app-acceptance} and~\ref{tab:app-dgroup}).}
\label{tab:gates}
\footnotesize
\setlength{\tabcolsep}{2.4pt}
\resizebox{\linewidth}{!}{\begin{tabular}{@{}lll@{}}
\toprule
\textbf{Gate} & \textbf{Threshold} & \textbf{TIDE (envelope)} \\
\midrule
\multicolumn{3}{@{}l}{\textit{Resolution and spectrum}} \\
resolution margin & $k_{\max}\eta\ge1.5$ & 1.55--4.8 \\
resolved dissipation & $\ge99.5\%$ & $99.84$--$100\%$ \\
dissipation-peak location & resolved & pass \\
spectral tail & strictly monotone & pass, no pile-up \\
\multicolumn{3}{@{}l}{\textit{Large scales and time}} \\
scale separation & $L_{\mathrm{box}}/L\ge4$ & $3.0$--$8.5$ (14/15)$^{\S}$ \\
advective step & $\mathrm{CFL}\le0.5$ & pass \\
dissipative step & $\Delta t/\tau_\eta\le0.05$ & pass \\
\multicolumn{3}{@{}l}{\textit{Stationarity and budget}} \\
energy drift & $\le1\%$ per $T_E$ & $\le0.66\%$ \\
injection--dissipation & closure within $2\%$ & $\le0.94\%$ (14/15) \\
sampling protocol & spin-up discarded & enforced \\
\multicolumn{3}{@{}l}{\textit{Symmetry and kinematics}} \\
component isotropy & cross $\le2\%$, comp.\ $\le5\%$, pooled & $0.9$--$1.8\%$ / $0.9$--$3.9\%$ \\
gradient isotropy & ratio near 1 & $1.002\pm0.010$ \\
incompressibility & $\le10^{-6}$ & $10^{-30}$--$10^{-28}$ \\
derivative skewness & $-S\in[0.45,0.60]$ & $0.515\pm0.015$ (forced) \\
\midrule
\multicolumn{3}{@{}l}{\textit{Equation-level residuals}} \\
divergence & $\le10^{-6}$ & $10^{-30}$--$10^{-27}$ \\
momentum residual & $\le10^{-2}$ & $\le5.8\times10^{-4}$ \\
step-halving ratio & $\in[2.5,6]$ (expect $\approx$4) & $3.96$--$4.00$ \\
pressure--Poisson consistency & machine precision & $\sim\!10^{-16}$ \\
\bottomrule
\end{tabular}}
\\[2pt]
{\footnotesize Isotropy, stationarity, and skewness-band gates apply to
the forced isotropic configurations; envelopes cover the configurations a
gate applies to. $^{\S}$The one configuration below the
scale-separation threshold is strong rotation ($L_{\mathrm{box}}/L=3.0$),
where the Taylor--Proudman quasi-two-dimensionalization grows the integral
scale; for that regime this item is reported rather than gated, as are
the isotropy items.}
\end{table}

Two conventions govern how these tables are read. First, the applicable
gates depend on the regime. Forced isotropic configurations are judged on
the full set of statistical gates (A1--A13), including component
isotropy as a hard gate. The
extended-physics configurations (rotation, passive scalar, stratification)
and the free-decay configurations are anisotropic or non-stationary
\emph{by construction}: Taylor--Proudman columns, scalar-gradient sheets,
buoyancy layering, and decaying energy are the physics under study, so
isotropy and stationarity gates do not apply there, and the hard gates
are resolution (including the Batchelor scale for the scalar case and the
Ozmidov scale with $Re_b$ for the stratified case), incompressibility, and
the equation-level D-group. Second, the reported values come from the
production acceptance runs; for three configurations they are quoted
from the release frame-set report instead, as noted in the tables.

\begin{table*}[t]\centering
\caption{Hard acceptance gates per configuration; dashes = not
applicable to the regime.}
\label{tab:app-acceptance}\footnotesize\setlength{\tabcolsep}{3.5pt}
\begin{tabular}{@{}lcccccc@{}}
\toprule
Config & $k_{\max}\eta$ & resolved diss.\ (A2) & drift (A7) & closure (A8) & $\langle(\nabla\!\cdot\!u)^2\rangle$ (A12) & Verdict \\
\midrule
$Re_\lambda 55$ & 3.003 & 100.000\% & 0.656\% & 0.04\% & $4.42\times10^{-29}$ & pass \\
$Re_\lambda 70$ & 2.125 & 99.989\% & 0.018\% & 0.68\% & $3.46\times10^{-29}$ & pass \\
$Re_\lambda 86$ & 1.614 & 99.881\% & 0.563\% & 0.66\% & $2.19\times10^{-29}$ & pass \\
$k_f{=}3$ & 1.629 & 99.903\% & 0.523\% & 0.94\% & $1.89\times10^{-29}$ & pass \\
$k_f{=}4$ & 1.689 & 99.931\% & 0.193\% & 0.20\% & $1.53\times10^{-29}$ & pass \\
$\tau{=}1$ & 1.836 & 99.957\% & 0.150\% & 8.03\%$^{\ddagger}$ & $2.42\times10^{-29}$ & pass \\
helical & 1.761 & 99.930\% & 0.007\% & 0.92\% & $3.48\times10^{-29}$ & pass \\
decay (hot-start) & 1.95--2.13 & --- & --- & --- & $\le2.21\times10^{-29}$ & pass \\
decay (Saffman) & 1.60 & --- & --- & --- & $2.84\times10^{-29}$ & pass \\
decay (Batchelor) & 1.55 & --- & --- & --- & $1.95\times10^{-29}$ & pass \\
decay (ABC) & 4.82 & --- & --- & --- & $6.34\times10^{-29}$ & pass \\
rotating (strong) & 2.10--2.86 & \multicolumn{5}{@{}p{0.62\textwidth}@{}}{frame-set gates all pass: Class~I; A2/A4, A12, D-group (with Coriolis source in D4).} \\
rotating (moderate) & Class~I & \multicolumn{5}{@{}p{0.62\textwidth}@{}}{frame-set gates all pass: Class~I; A2/A4, A12, D-group.} \\
stratified & 1.853 & \multicolumn{5}{@{}p{0.62\textwidth}@{}}{frame-set gates pass: Class~I; A2 99.95\%, $L_{\mathrm{box}}/L=5.4$, $Re_b=39.2$, Ozmidov scale resolved ($\eta<l_O<L$); drift and isotropy gates do not apply (anisotropic).} \\
passive scalar & 1.63 & \multicolumn{5}{@{}p{0.62\textwidth}@{}}{frame-set gates pass: Class~I with Batchelor-scale margin $k_{\max}\eta_B=1.55$; A2 99.84\%, A12 $4.4\times10^{-30}$, D3 ratio 3.99, D4 $1.6\times10^{-16}$.} \\
\bottomrule
\end{tabular}
\\[2pt]{\footnotesize $^{\ddagger}$$8.0\%$ over its acceptance window, a
short-window effect; reported, not suppressed.}
\end{table*}

\begin{table*}[t]\centering
\caption{D-group residuals per configuration (protocol in
Section~\ref{sec:acceptance}).}
\label{tab:app-dgroup}\footnotesize\setlength{\tabcolsep}{4pt}
\begin{tabular}{@{}lcccc@{}}
\toprule
Config & D1 divergence & D2 momentum & D3 order & D4 pressure--Poisson \\
\midrule
$Re_\lambda 86$ & $5.21\times10^{-30}$ & $5.76\times10^{-4}$ & 3.96 & $1.8\times10^{-16}$ \\
$Re_\lambda 70$ & $5.60\times10^{-30}$ & $2.67\times10^{-4}$ & 3.98 & $1.6\times10^{-16}$ \\
$Re_\lambda 55$ & $4.37\times10^{-30}$ & $1.4\times10^{-4}$ & 3.98 & $1.9\times10^{-16}$ \\
$k_f{=}3$ & $3.96\times10^{-30}$ & $4.46\times10^{-4}$ & 3.97 & $1.8\times10^{-16}$ \\
$k_f{=}4$ & $3.61\times10^{-30}$ & $3.94\times10^{-4}$ & 3.96 & $1.7\times10^{-16}$ \\
$\tau{=}1$ & $5.21\times10^{-30}$ & $4.37\times10^{-4}$ & 3.97 & $2.0\times10^{-16}$ \\
passive scalar & $4.4\times10^{-30}$ & $2.68\times10^{-5}$ & 3.99 & $1.6\times10^{-16}$ \\
rotating (moderate) & $7.4\times10^{-30}$ & $1.16\times10^{-4}$ & 3.99 & $9.3\times10^{-17}$ \\
helical & $6.56\times10^{-30}$ & $3.64\times10^{-4}$ & 3.97 & $1.8\times10^{-16}$ \\
decay (hot-start) & $1.81\times10^{-27}$ & $5.41\times10^{-8}$ & 4.00 & $1.3\times10^{-16}$ \\
decay (Saffman) & $4.98\times10^{-29}$ & $4.11\times10^{-6}$ & 3.99 & $1.2\times10^{-16}$ \\
decay (Batchelor) & $7.12\times10^{-29}$ & $2.59\times10^{-6}$ & 4.00 & $1.2\times10^{-16}$ \\
decay (ABC) & $1.52\times10^{-29}$ & $7.10\times10^{-7}$ & 4.00 & $1.7\times10^{-16}$ \\
\bottomrule
\end{tabular}
\\[2pt]{\footnotesize Configurations not listed here have their D-group verification reported in the released per-case files rather than reproduced in this table.}
\end{table*}

\begin{figure*}[t]
  \centering
  \includegraphics[width=\linewidth]{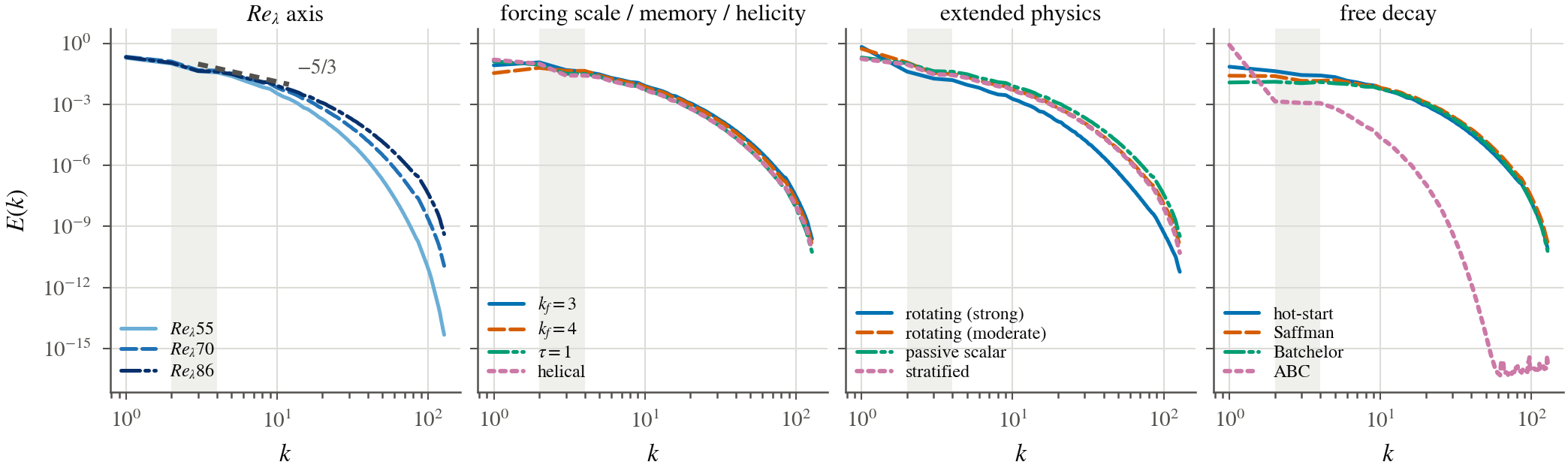}
  \caption{Energy spectra of all 15 configurations (forcing band shaded,
  $-5/3$ reference); all fall monotonically to the resolution scale; the
quasi-steady ABC tail bottoms out at the fp64 roundoff floor,
$\sim\!10^{-16}$.}
  \label{fig:spectra}
\end{figure*}

\section{D-group: Equation-Level Verification Protocol}
\label{app:dgroup}
The D-group verifies the released fields against the governing equations.
D1 (divergence) and D4 (velocity--pressure consistency) are algebraic
checks on single frames: D1 evaluates
$\langle(\nabla\!\cdot\!\mathbf u)^2\rangle/\langle|\nabla\mathbf u|^2\rangle$
spectrally; D4 solves $\nabla^2 p=-\partial_i\partial_j(u_iu_j)$ (with
Coriolis/buoyancy source terms for the extended physics) and verifies both
the Poisson residual and the balance of $\nabla p$ against the irrotational
part of the advection term. D2 (momentum residual) uses dedicated frame
triplets $(u^{n-1},u^n,u^{n+1})$ exported with the OU forcing state frozen
between frames, so that the centered-difference time derivative can be
compared against $\mathcal P(u\times\omega)+\nu\nabla^2 u+f$ with no
stochastic contamination; the residual then contains only the $O(h^2)$
time-truncation error, which D3 confirms by halving the step and checking
the residual ratio falls in $[2.5,6]$ (measured $\approx4$). The frozen
protocol matters: with live stochastic forcing the apparent residual is
inflated $10$--$30\times$ and would mask genuine equation errors.

\section{Corpus Reference Tables}
\label{app:reftables}
Table~\ref{tab:decay} tabulates the free-decay family behind the
initial-condition axis of Section~\ref{sec:overview}; the column to
read is the energy fold over the released window, which shows the four
initial states decaying by visibly different amounts. The
per-configuration parameter table (Table~\ref{tab:configs}) and the
landscape comparison (Table~\ref{tab:related}) are in the main text.

\begin{table}[t]
\centering
\caption{The four free-decay families; $K$ fall = kinetic-energy drop
over the released window, whose length is given in initial
eddy-turnover times $\tau_0$.}
\label{tab:decay}
\small
\setlength{\tabcolsep}{5pt}
\resizebox{\linewidth}{!}{\begin{tabular}{@{}llcc@{}}
\toprule
\textbf{Family} & \textbf{Initial condition} & \textbf{$K$ fall} & \textbf{Window} \\
\midrule
Saffman   & spectrum $p{=}2$          & $24\times$ & $4.7\,\tau_0$ \\
Batchelor & spectrum $p{=}4$          & $35\times$ & $5.2\,\tau_0$ \\
hot-start & matured from forced state & $6\times$  & $1.7\,\tau_0$ \\
ABC       & maximal-helicity Beltrami & $55\times$ & $5.6\,\tau_0$ \\
\bottomrule
\end{tabular}}
\end{table}

\section{Per-Category Slice Gallery}
\label{app:gallery}
\begin{figure}[t]
  \centering
  \includegraphics[width=\linewidth]{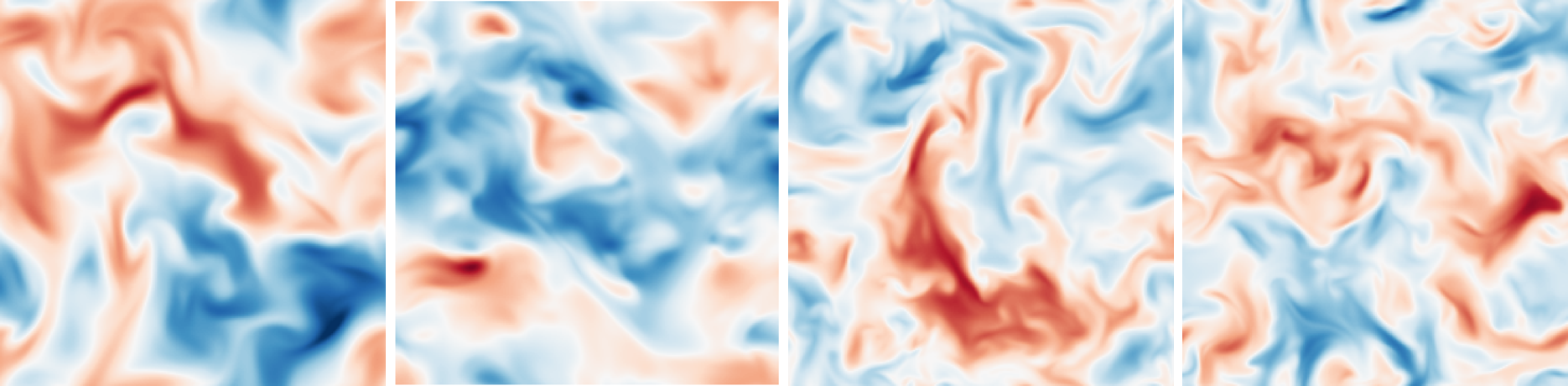}
  \caption{Forced HIT, $u_x$ mid-plane slices:
  $Re_\lambda\in\{55,70,86\}$ and $k_f{=}4$ (shared colormap).}
  \label{fig:cat-forced}
\end{figure}

\begin{figure}[t]
  \centering
  \includegraphics[width=\linewidth]{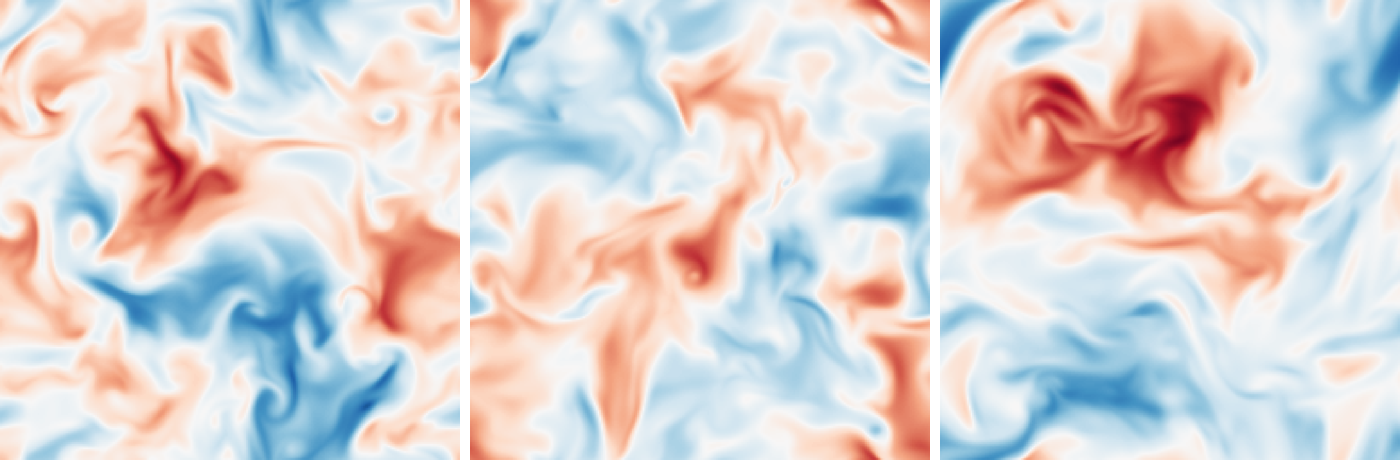}
  \caption{Remaining forced configurations: $k_f{=}3$, $\tau{=}1$,
  helical.}
  \label{fig:more-forced}
\end{figure}

\begin{figure}[t]
  \centering
  \includegraphics[width=\linewidth]{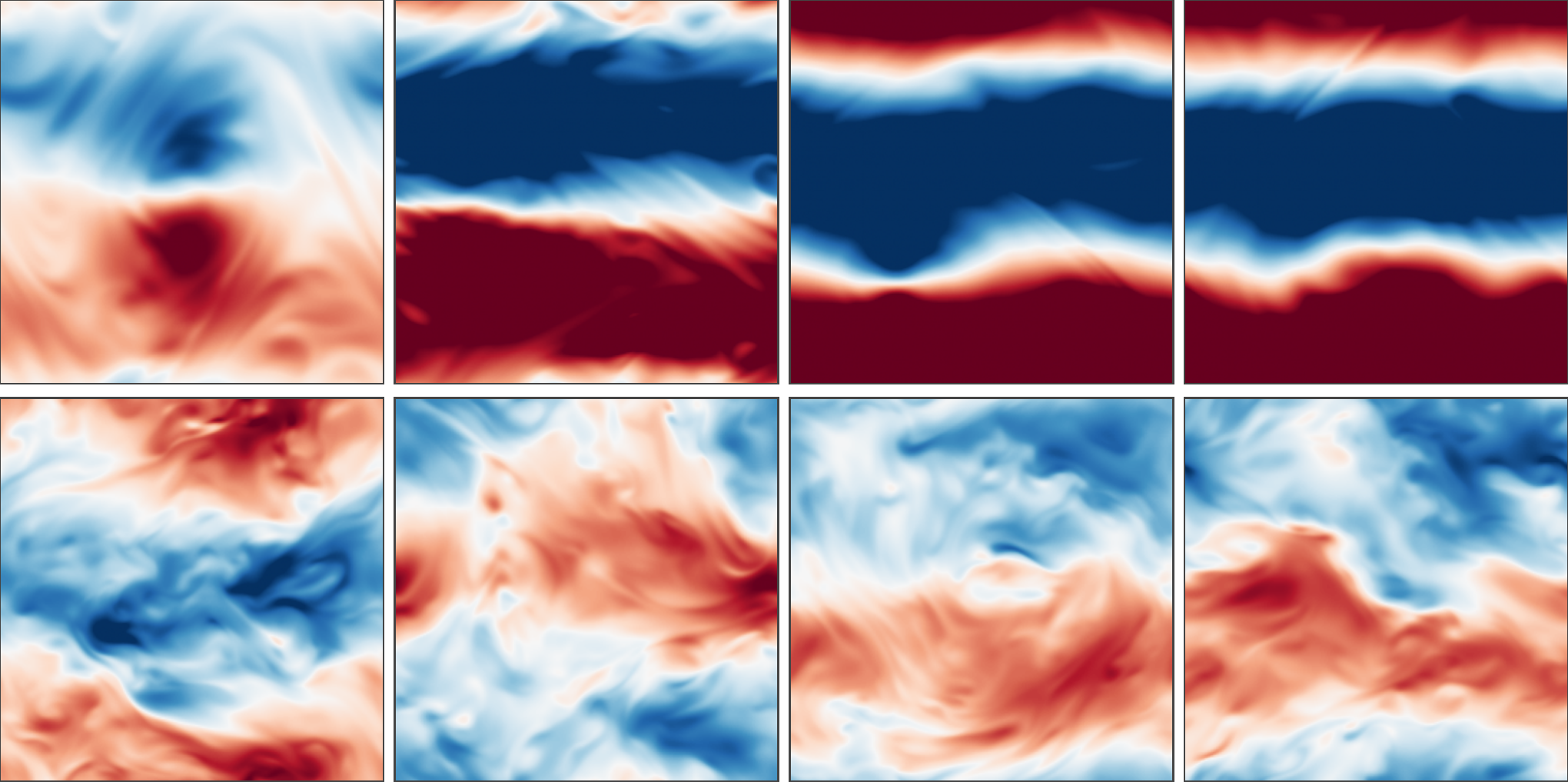}
  \caption{Rotating turbulence, $\omega_z$ on a vertical slice at four
  times: strong (top) and moderate (bottom) rotation.}
  \label{fig:cat-rotating}
\end{figure}

\begin{figure}[t]
  \centering
  \includegraphics[width=\linewidth]{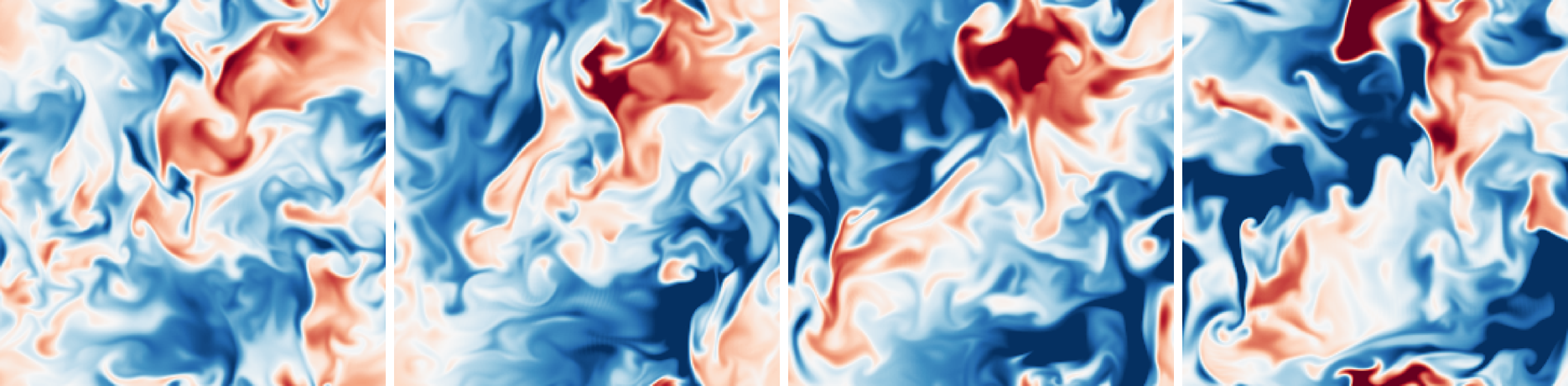}
  \caption{Passive scalar $\theta$ ($Sc{=}1$) at four instants.}
  \label{fig:cat-scalar}
\end{figure}

\begin{figure}[t]
  \centering
  \includegraphics[width=\linewidth]{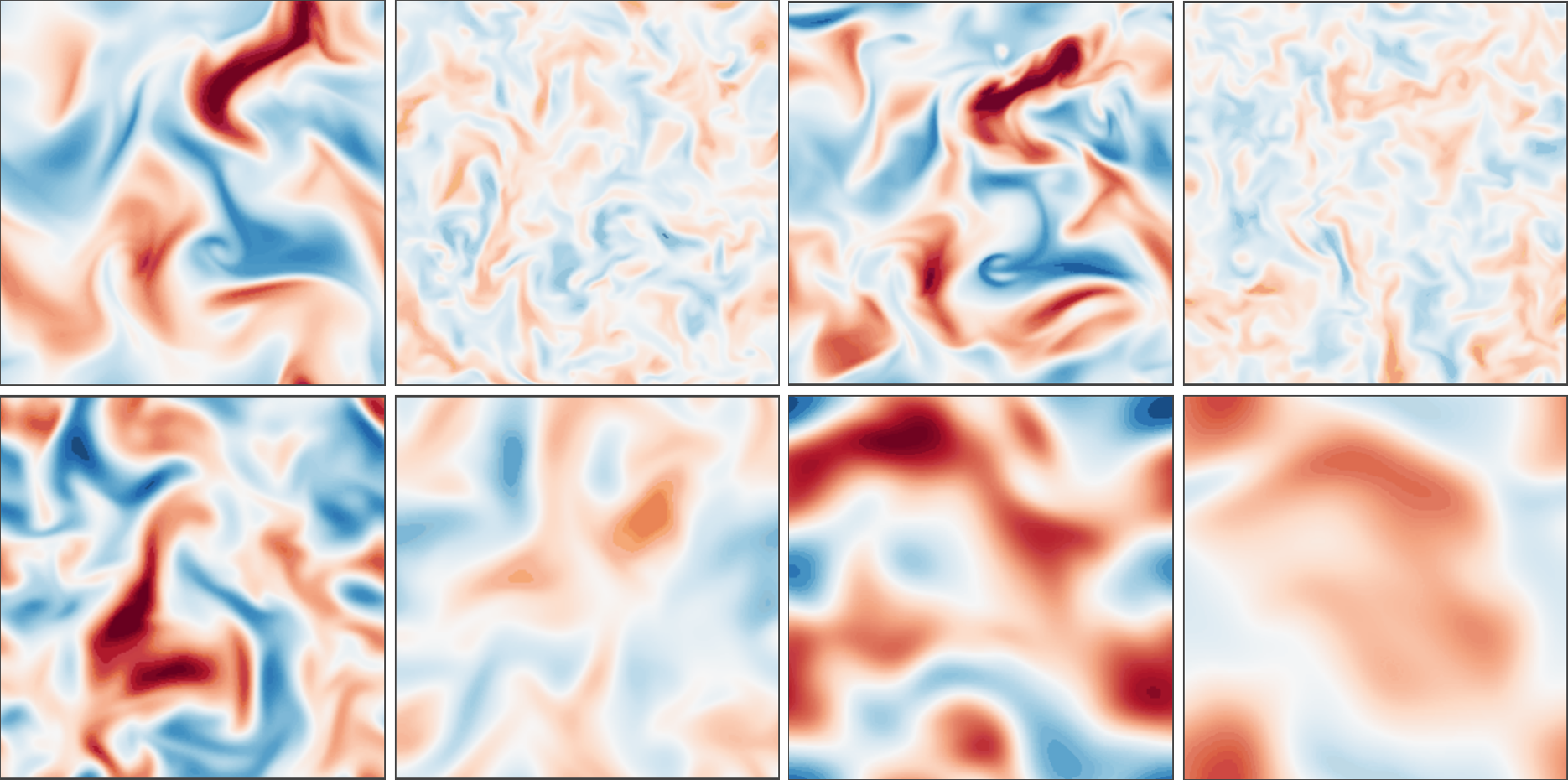}
  \caption{Free-decay family, an early and a later frame per case.}
  \label{fig:cat-decay}
\end{figure}

\begin{figure}[t]
  \centering
  \includegraphics[width=\linewidth]{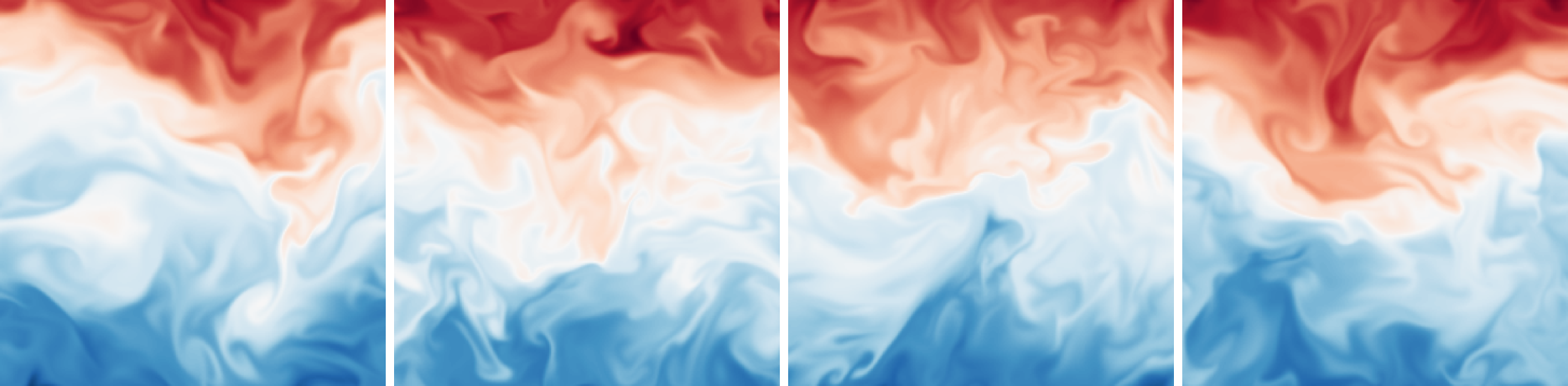}
  \caption{Stratified turbulence: buoyancy $b$ on a vertical plane at
  four times.}
  \label{fig:cat-stratified}
\end{figure}

Figure~\ref{fig:taxonomy} in the main text shows representative slices
per regime family. The per-category strips below give every configuration and multiple
frames behind them, each caption stating the quantity shown; they are the visual counterpart
of the diversity claim of Section~\ref{sec:overview}, and the pattern
to check is that configurations differ visibly across a strip while
frames within one configuration remain statistically alike.
Figures~\ref{fig:cat-forced} and~\ref{fig:more-forced} cover the forced
configurations; Figures~\ref{fig:cat-rotating}, \ref{fig:cat-scalar},
\ref{fig:cat-decay}, and~\ref{fig:cat-stratified}
cover the rotating, scalar, decaying, and stratified families in turn.

\section{Data Format and Access}
\label{app:format}
Each configuration ships as one chunked zarr store (velocity
$[N,3,256^3]$, pressure $[N,256^3]$, optional $\theta/b$, per-frame $t$
and $k_{\max}\eta$, seed index; one frame per chunk, zstd), together with
frozen normalizer JSONs, split manifests, and loader examples.
Computation is fp64 and storage fp32: in
fp32 computation the truncation scale accumulates energy in the highest
wavenumber shells until the field collapses, whereas fp64 runs show no
such accumulation over the full trajectory lengths run; fp32 storage was
verified by checking that the incompressibility residual of the
round-tripped field stays far inside the acceptance gate
($\approx\!10^{-13}$ against $10^{-6}$). Ensemble independence is
machine-verified per release: every seed's checkpoint checksum is
unique, pairwise first-frame differences are at field scale, and the
release-blocking independence test passes with no exemptions. The test is release-blocking and has previously rejected a candidate
release. The corpus is hosted on
HuggingFace, one repository per configuration so that a single regime can
be fetched without the whole release, indexed from a landing repository
that also carries the datasheet~\cite{gebru2021datasheets}, the split
manifest, and the per-configuration manifests; Croissant metadata is
served for each repository. A versioned DOI record archives the
datasheet, the split manifest, and a snapshot of the generation,
acceptance, and benchmark code, and is the citable anchor for the
release. Data are under CC-BY-4.0 and code under MIT; errata and
additions are released as new versions, and a maintenance contact ships
with the datasheet.

\section{Baseline Architectures and Hyperparameters}
\label{app:baselines}

\begin{table}[t]
\centering
\caption{The five learned baselines at matched capacity, with the
trivial and informed references.}
\label{tab:baselines}
\footnotesize
\setlength{\tabcolsep}{3.5pt}
\begin{tabular}{@{}lllc@{}}
\toprule
\textbf{Model} & \textbf{Native-$256^3$ input handling} & \textbf{Params} & \textbf{Ref.} \\
\midrule
\multicolumn{4}{@{}l}{\textit{Learned baselines (capacity-matched)}} \\
FNO3d      & full field, gradient checkpointing  & 28.3\,M & \cite{li2021fno} \\
TFNO & full field, gradient checkpointing  & 30.3\,M & \cite{kossaifi2023tfno} \\
Spectral U-Net & $128^3$-crop training, native eval & 28.2\,M & \cite{wen2022ufno} \\
Transolver & full field, patch tokenization       & 29.7\,M & \cite{wu2024transolver} \\
DeepONet-3D & full field, chunked trunk            & 7.0\,M$^{\ast}$ & \cite{lu2021deeponet} \\
\midrule
\multicolumn{4}{@{}l}{\textit{Trivial references (no learning, no equations)}} \\
persistence      & lag-1 (rollout)          & --- & --- \\
spectral interp. & spectral upsample (super-res) & --- & --- \\
identity         & masked field (recon)     & --- & --- \\
dyn.\ Smagorinsky & LES closure (subgrid stress) & --- & \cite{pope2000turbulent} \\
\midrule
\multicolumn{4}{@{}l}{\textit{Informed references (given the governing equations)}} \\
Poisson solve    & spectral $u\rightarrow p$ (pressure) & --- & --- \\
equations-informed integrator & spectral solve (forecasting) & --- & --- \\
\bottomrule
\end{tabular}
\\[2pt]
{\footnotesize $^{\ast}$memory ceiling of its dense trunk. TFNO and
Spectral U-Net are audited variants (Appendix~\ref{app:deviations}).}
\end{table}

The five learned baselines are matched by capacity to $28$--$30.3$\,M
parameters (FNO3d $28.3$, TFNO $30.3$, Spectral U-Net $28.2$,
Transolver $29.7$), with DeepONet-3D at its architectural ceiling of
$7.0$\,M, since its dense trunk over the full grid exhausts a 32\,GB GPU at
the next capacity step. Matching is done upward, by enlarging the smaller
models, not by shrinking the larger ones. TFNO uses Tucker rank 18 with $18^3$ retained Fourier modes per block,
against $12^3$ for FNO3d and the Spectral U-Net, $3.4$ times as many
retained modes. Parameter count alone does not
measure capacity for this model: at essentially identical counts
($\approx\!30.3$\,M), halving the spectral width degrades a controlled
overfitting probe by a factor of $6.4$ and, in a run outside the
reference matrix, pushed rollout past the persistence level
on the lowest-Reynolds configuration. Capacity matching is therefore
audited by the probe, which is part of the released code, rather than by
the parameter count.

Evaluation is native $256^3$ for every model, and so is training except
where noted below
(Section~\ref{sec:native-eval}). Each model meets the memory cost in
its implementation layer: the spectral operators recompute activations by
gradient checkpointing, Transolver tokenizes the field into patches,
DeepONet-3D chunks its trunk evaluation, and the Spectral U-Net alone
trains on $128^3$ crops because a full-field backward pass structurally
exceeds memory, while still being evaluated natively. Training uses AdamW (learning rate
$10^{-3}$, weight decay $10^{-5}$) with gradient-norm clipping at 1.0
applied after accumulation, and a cosine schedule stepped once per epoch
that anneals the learning rate to zero over 25 epochs, so completing 25
epochs is the normal end of training; early stopping (patience 8) on the single-step validation loss ends
some runs before epoch 25, and evaluation uses the checkpoint with the
lowest single-step validation loss. The learning rate is identical across
baselines by design (Appendix~\ref{app:insights} records why a lower rate
was rejected), and all models train from the same reference seed governing
initialization and shuffling, with up to four replicate seeds per forecasting
cell (Table~\ref{tab:app-seeds}); these seeds are distinct from the DNS
trajectory identifiers used by the split. Rollout uses residual prediction. All
optimizer settings ship with the released training configuration files.

\begin{table}[t]
\centering
\caption{The benchmark tasks: single-step maps
$f:(C_{\mathrm{in}},256^3)\rightarrow(C_{\mathrm{out}},256^3)$; AR =
autoregressive rollout.}
\label{tab:tasks}
\footnotesize
\setlength{\tabcolsep}{2.5pt}
\resizebox{\linewidth}{!}{\begin{tabular}{@{}llccl@{}}
\toprule
\textbf{Task} & \textbf{Pair} & $C_{\mathrm{in}}$ & $C_{\mathrm{out}}$ & \textbf{Main metric} \\
\midrule
P1 Forecasting   & $u(t)\rightarrow u(t{+}\Delta t)$; AR-20 eval & 4 & 4 & nRMSE, enstrophy ratio \\
P2 Super-res.\  & $D_s(X)$ ($\times4$) $\rightarrow X$      & 4 & 4 & nRMSE, spectrum-$L_2$ \\
P3 Sparse recon.\ & $M{\odot}X$ (95\% miss) $\rightarrow X$  & 4 & 4 & nRMSE \\
P4 Pressure recovery & $(u,v,w)\rightarrow p$                    & 3 & 1 & nRMSE, Poisson res.\ \\
P5 Subgrid stress & $\bar G * X \rightarrow \tau_{ij}^{sgs}$ & 4 & 6 & $\tau_{ij}$ corr., backscatter \\
P6 Denoise\,$^\dagger$ & $X{+}\eta \rightarrow X$            & 4 & 4 & nRMSE, high-band \\
P7 Inversion\,$^\dagger$ & $X \rightarrow (Re_\lambda,\nu)$  & 4 & --- & MAE \\
\bottomrule
\end{tabular}}
\\[2pt]
{\footnotesize $^\dagger$specified in the appendix, not in the reference
results.}
\end{table}

\begin{table}[t]
\centering
\caption{Generalization axes: each transfer's test configurations are
held out of its training split.}
\label{tab:ggroup}
\footnotesize
\setlength{\tabcolsep}{3pt}
\resizebox{\linewidth}{!}{\begin{tabular}{@{}lll@{}}
\toprule
\textbf{Axis} & \textbf{Train $\rightarrow$ Test} & \textbf{Probes} \\
\midrule
G1 cross-$Re$        & $Re_\lambda 55, 70\rightarrow 86$$^{\S}$ & extrap.\ to finer scales \\
G2 cross-physics     & passive scalar $\rightarrow$ stratified & unseen Boussinesq physics \\
G3 cross-forcing     & $k_f 2, 3\rightarrow 4$$^{\S}$          & forcing-regime transfer \\
G4 forced$\rightarrow$decay$^{\dagger\dagger}$ & steady $\rightarrow$ free decay  & drive removed (unconditioned) \\
G5 cross-phys.\ type & rotating $\leftrightarrow$ isotropic$^{\ast}$  & both directions scored \\
\bottomrule
\end{tabular}}
\\[2pt]
{\footnotesize $^{\S}$Scored per source; G1 also with a joint
model (Appendix~\ref{app:joint}). $^{\ast}$G5 runs on the rotating configurations only.
$^{\dagger\dagger}$G4 is evaluated within each flow
(Section~\ref{sec:benchmark}).}
\end{table}

\section{Protocol Details}
\label{app:protocol}
The shared recipe of Section~\ref{sec:native-eval} rests on measured
evidence. Under absolute-field regression we measured the baselines
collapsing toward a near-mean output, scoring worse than persistence,
which is what fixed residual prediction; the learning-rate study
(Appendix~\ref{app:recipe}) shows that the sensitivity to the shared
$10^{-3}$ is mild in nRMSE but not in the enstrophy ratio. The voxel-defined effective batch keeps batch size comparable
across resolutions and keeps the per-epoch schedule from being starved
of steps; a reproducer changing batch size or resolution should account
for the resulting change in optimizer steps per epoch.

How each architecture meets the full-field memory cost is described in
Appendix~\ref{app:baselines}. The Spectral U-Net's cropped training is an
acknowledged asymmetry in the comparability invariant, and
Appendix~\ref{app:insights} records why cropped training withholds part
of the forced scales in our configurations.

\section{Benchmark Slice: Quality Score}
\label{app:slice}
Trajectories are ranked within each configuration by a model-independent
quality score built from standardized per-trajectory physical
quantities: resolution margin (instantaneous $k_{\max}\eta$ above 1.5),
stationarity drift (energy drift per $T_E$), the number of timeline
splices left by resumed runs, and the frame count, equally weighted.
Isotropy is assessed by pooling over a configuration's trajectories, so
it cannot rank them and does not enter the score, and the drift term
enters where the per-trajectory statistic is recorded (9 of the 15
configurations); the terms used for each configuration ship with the
slice manifest so the ranking can be recomputed. Ranking by descending
score, the top three trajectories form the training split, the fourth is
validation, and the next three are test, with two exceptions recorded in
the manifest: the flagship tests on all seven remaining trajectories, and
the stratified configuration has no training split and serves only as a
transfer target. The three splits are disjoint, verified at the level of
stored rows rather than trajectory labels, and the trajectories outside
the slice remain available for training extensions
(Appendix~\ref{app:recipe}). The slice is generated by the released
\texttt{make\_benchmark\_slice.py}; hand-edited manifests are rejected,
since the selection rule is itself a public claim. Every baseline reads
the same manifest and the same frozen normalizer, fit on that
configuration's training trajectories. Because the highest-scoring
trajectories train, the test sets differ physically from the training
sets: on $Re_\lambda 86$ the test trajectories have $Re_\lambda$ higher by
$3.2\%$, dissipation higher by $16.9\%$, and $k_{\max}\eta$ lower by
$3.8\%$, and across the configurations with measured statistics the
test-set dissipation is higher by $0.2$--$24.7\%$ and $Re_\lambda$
differs by $-11.2\%$ to $+5.0\%$. Appendix~\ref{app:splits}
measures what this shift does to the benchmark.

The nominal test count overstates
independence: with a measured decorrelation time of $3.3\,T_L$, the
effective sample size on the 51-window forced records is
$N_{\mathrm{eff}}\approx9$, essentially invariant to the evaluation
stride, because the independent information is set by trajectory length
times trajectory count and sparser sampling only discards samples.
Sampling uncertainty should be read against this count rather than the
nominal one; evaluation windows start every 8 frames, and the nominal counts are
51 windows on the standard forced records (on the flagship, whose seven
test trajectories are longer, these cover about two trajectories, and
all 135 are scored in Table~\ref{tab:app-seeds}), 37 on the scalar
record, 12 on the hot-start and ABC decays, and 6 on the Saffman and
Batchelor decays, and results on the shorter
records carry proportionally larger uncertainty. In training, by contrast,
window overlap is standard augmentation and independence carries no
meaning; the reference stride of 4 frames trades data
volume against the strong correlation of consecutive frames, and denser
sampling is evaluated in Appendix~\ref{app:recipe}.

\section{Metric Definitions}
\label{app:metrics}
The metric suite comprises nRMSE, spectral errors split by wavenumber
band, spectrum-$L_2$, effective prediction time, enstrophy ratio,
vorticity-PDF tails, the P4 pressure--Poisson residual, and the P5
stress correlation and backscatter fraction; full formulas ship with the
released documentation. The high-band spectral error, like the
Poisson residual and the subgrid energy transfer, are absolute
quantities computed after de-normalization, so they carry physical units
and remain comparable across configurations. Band errors sum the
half-spectrum without Hermitian weights. Implementations are validated at two
levels: known-answer synthetic fields check each metric's algebra, and
corpus-level tests with mutation coverage check every metric and
non-learning reference against physical invariants of the released
fields, for the reasons recorded in Appendix~\ref{app:insights}. The
released result files carry further measured quantities beyond the
tabulated axes, including high-band spectral errors, a spectral
centroid drift, and vorticity-tail statistics.

\section{Baseline Architecture Deviations}
\label{app:deviations}
Every baseline was audited against its published implementation, not only
the ones that failed to train, and each deviation is recorded with its
reason. Table~\ref{tab:deviations} summarizes the audit. Two models
depart from their published counterparts in structure: TFNO applies an
independent Tucker factorization per spectral corner rather than the
shared factorization of the reference, and the Spectral U-Net is a U-Net
encoder--decoder pyramid with spectral blocks rather than the per-layer
bypass of U-FNO~\cite{wen2022ufno}. Results for these two should be read as results for the
audited variants specified here, not for the original implementations.

\begin{table}[t]
\centering
\caption{Audited deviations from the published implementations; two
models are renamed accordingly.}
\label{tab:deviations}
\footnotesize
\setlength{\tabcolsep}{4pt}
\begin{tabular}{@{}lp{0.62\linewidth}@{}}
\toprule
\textbf{Model} & \textbf{Principal deviations (full list in the released audit)} \\
\midrule
FNO3d & Core faithful to Li et al.; no coordinate channels (deliberate, periodic homogeneous
flow has no absolute position), two-layer lifting/projection, gradient checkpointing. \\
TFNO & Mathematically a low-rank FNO3d; per-corner independent Tucker cores (no cross-corner
or cross-layer sharing), real factor matrices, Tucker rank 18, equal to the retained mode count ($18^3$ modes per block
against $12^3$ for FNO3d). \\
Spectral U-Net & A U-Net encoder--decoder pyramid with spectral blocks, not the per-layer U-Net
bypass of U-FNO~\cite{wen2022ufno}; shares the ``spectral $\times$ multiscale'' design space but a different
topology. \\
Transolver & One missing weighted-mean normalization step, present in the official code, restored
(without it the model did not train); LayerScale and patch-16 tokenization recorded as deviations. \\
DeepONet-3D & Convolutional branch and grid trunk rather than the sensor-MLP branch of the
original; a 3D variant. \\
\bottomrule
\end{tabular}
\end{table}

\section{Discussion and Limitations}
\label{app:discussion}

\paragraph{The Reynolds ceiling is a resolution consequence.}
At a fixed $256^3$ grid, higher $Re$ shrinks the Kolmogorov scale until
the Class~I requirement $k_{\max}\eta\ge1.5$ fails; two retained boundary
anchors measure where, with the resolved-dissipation fraction reading
$99.47\%$ at $Re_\lambda=98$ and $98.3\%$ at $Re_\lambda=111$ against a
$99.5\%$ gate, placing the ceiling at $Re_\lambda\approx86$--$90$. The
ceiling applies to the isotropic cascade: rotation and the Beltrami
initial state suppress the forward cascade, so those configurations
reach nominally higher $Re_\lambda$ (Table~\ref{tab:configs}) at the
same resolution margin. TIDE
therefore does not compete with high-$Re$ databases on turbulence
intensity~\cite{li2008jhtdb}; its contribution is structure, not
Reynolds number.

\paragraph{What the OOD axes do and do not claim.}
Relative to the reference configuration, the forcing-scale, rotation,
scalar, and stratification settings each change one parameter; the
helical setting changes two (forcing type and forcing variance); on the
Reynolds axis only the step between $Re_\lambda 70$ and $86$ changes
viscosity alone, since $Re_\lambda 55$ also differs in forcing variance
and, relative to $Re_\lambda 86$, in frame spacing ($+12.7\%$); the decay
configurations change viscosity and initial condition. The hot-start
decay branches from the forced $Re_\lambda 86$ trajectories at
$t{=}40$, so the two configurations are not independent, and its frame
spacing is $16\%$ larger. We therefore report G1, G2, G3, and G5 as
regime transfers with interpretable direction rather than single-factor
attributions, and do not score G4 as a transfer at all. The
forcing-memory axis has two points.

\paragraph{Known biases of the released frames.}
The per-frame resolution gate drops dissipation-peak instants, so extreme
intermittency is under-sampled and high-order gradient statistics are
lower bounds; the gate removes frames in four of the 15
configurations ($19.1\%$ of the $Re_\lambda 86$ frames, $32$--$33\%$ for
Saffman, $33\%$ for Batchelor, and about $9\%$ for the scalar, estimated
from frame counts), and on $Re_\lambda 86$ the removed instants carry at
least $1.28$ times the retained mean dissipation. Removed frames are not
in the corpus and cannot be scored.
Free-decay exponents are reported for reference only, since
the fitted rate depends strongly on the virtual origin; decay cases
should be judged by equation-level consistency and their decay dynamics
rather than by matching textbook exponents. Storage is fp32 while computation
is fp64, so bit-exact regeneration is hardware-bound while statistical
reproduction is not. The frozen normalization constants are fit on each
configuration's training trajectories only, and the released sidecar
records the fitting scope, so the choice is auditable rather than
asserted. One consequence follows from that choice: the scale is the
training-split extremum, so held-out frames may fall marginally outside
the nominal range, which is correct behavior rather than a defect.
Studies extending the corpus should refit the constants on their own
training split.

\paragraph{Scope of the numerics and of the benchmark.}
All data come from one pseudo-spectral solver family on a periodic box,
so TIDE is a turbulence-physics corpus rather than a
cross-discretization benchmark; wall-bounded flows, compressible
effects, and multi-solver comparisons
are out of scope. The reference matrix is not uniform: the six
main-table configurations carry every model a task admits, four on
forecasting and three or four on each single-frame task, the hot-start, Saffman,
and both rotating configurations carry three learned models and a
non-learned reference per single-frame task, and the remaining extended
configurations carry a reduced model set, stated per table. Eight
single-frame models on $k_f{=}3$, marked in Tables~\ref{tab:app-superres}, \ref{tab:app-recon},
\ref{tab:app-pressure}, and~\ref{tab:app-sgs},
were trained at frame stride 1.
Every empty cell has one of the following causes. Transolver has no
pressure or subgrid-stress entries because its output head has as many
channels as its input and cannot emit the one-channel pressure or the
six-channel stress. The stratified configuration has no training split
(all eight seeds are test seeds), so it has no in-distribution cells and
serves only as a transfer target. Four-channel and five-channel models
cannot be transferred into each other because the input layer is built
for the training channel count. DeepONet-3D is benchmarked on pressure
only. The remaining empty cells lie outside the reference matrix: Spectral
U-Net pressure,
TFNO reconstruction and pressure on helical, ABC, and Batchelor, the
passive-scalar single-frame cells of TFNO, the Spectral U-Net, and
DeepONet-3D, the TFNO and Spectral U-Net transfers from moderate rotation
and from the passive scalar, and Transolver super-resolution and FNO3d
reconstruction on the extended configurations.
Generative surrogates (diffusion models and other iterative-refinement
samplers) are scoped out of the reference matrix: candidate
configurations were prepared, but their multi-step sampling protocols
require evaluation choices our single-forward protocol does not fix, so
we state the boundary rather than score them under a protocol not
designed for them. The protocol answers one question, whether
a single-step operator learned on one regime reproduces the dynamics
zero-shot; a model designed to train on rollouts, consume a history, or
be corrected by a solver in the loop is measured outside its intended
setting, and a quasi-steady flow degrades the forecasting question
itself, as the ABC configuration shows. Forecasting results are medians over five
training seeds against an effective sample size
$N_{\mathrm{eff}}\approx9$, and orderings are reported as separable or
not rather than by the smallest value.

\paragraph{Ethics and compute.}
TIDE is synthetic simulation data of a canonical physical system, with no
human subjects, personal data, or scraped content, and misuse potential is
low. The main societal cost is compute: corpus production is estimated at
approximately 105 GPU-hours (80--130 under the spin-up uncertainty stated
in Appendix~\ref{app:compute}), extrapolated from measured per-step
timing, and the reference benchmark at approximately 97, roughly 200
GPU-hours in total on consumer GPUs. The corpus and tooling are released so this compute need not be
repeated.

\paragraph{Future directions.}
The generalization results separate
two candidate causes of transfer error: training coverage, measured
through joint training across regimes (Appendix~\ref{app:joint}), and
the missing forcing state, measured through a short history and an explicit forcing channel, together with
the integrator bound on what the current forcing state can contribute
(Appendix~\ref{app:conditioning}). Neither closes the gaps on its own.
Cross-regime pretraining is feasible with existing architectures but is
a study in its own right; an operator trained across many physical
regimes and conditioned on which regime it is in would have the defining
ingredients of a foundation model for 3D incompressible turbulence, and
we regard such pretraining as a continuation of these results rather
than a separate ambition, with TIDE as its platform. Additional physics
axes under the same acceptance discipline, rollout-aware training, and
community metric extensions are the other open directions.

\section{Design Insights}
\label{app:insights}
The corpus and the benchmark protocol were each backed by a measurement. We record the findings, each stated as
the observation and the measurement behind it; whether they hold beyond
this corpus is untested.

\paragraph{Deterministic forcing is metastable at this scale.}
Deterministic band forcing, in both fixed-power and energy-preserving
variants, can remain stationary for over a hundred eddy-turnover times
and then destabilize, through runaway energy in the lowest forced shell
or through relaminarization, so a short validation window produces a
false positive. Random-phase OU forcing destroys this metastable
attractor, which is its original design
motivation~\cite{eswaran1988forcing}, and the production runs show no
collapse over the full trajectory length. The negative result is
documented in the released repository.

\paragraph{Cropped training withholds the forced scales in our configurations.}
The forcing acts at $k=2$--$4$ across the corpus, so the energy-containing
scale is the box scale, and a $128^3$ sub-block of the $256^3$ periodic box
cannot represent modes below $k=2$. Training on cropped inputs therefore withholds
information about the large-scale dynamics that the task depends on,
which is why the protocol operates on the full field wherever memory
allows. The Spectral U-Net is the exception: it sees the same voxels and
optimizer steps per epoch as the other models, as eight deterministic
$128^3$ crops. Trained on the full field instead (one seed), it is
worse on $Re_\lambda 70$ ($+0.069$ [$+0.063$, $+0.074$], lower on 0/3
trajectories) and not separable from its cropped seeds on
$Re_\lambda 86$, where its enstrophy ratio reaches 361.

\paragraph{Our tiled evaluation variant rewarded inaction.}
We ablated a tiled-inference variant in which the field is cut into
sub-blocks and predictions reassembled. Its reassembly seams inject
high-wavenumber discontinuities that autoregression amplifies: over a
20-step rollout the enstrophy grew to $19$--$30\times$ the true value
(Appendix~\ref{app:recipe}). Crucially the seam is invisible on a collapsed model whose output
equals its (continuous) input, so a tiled pipeline scores a do-nothing
model as seamless and systematically rewards inaction. This is the direct
reason our protocol evaluates natively, and should be expected in other pipelines that reassemble predictions from sub-blocks.

\paragraph{Training-time comparisons are read against optimizer steps.}
Models can remain in a near-persistence output regime for several
hundred optimizer steps under this recipe, over which window validation
loss is insensitive to the learning rate; comparisons are therefore read
against optimizer-step counts. Under the final protocol the
learning-rate sensitivity is mild in nRMSE (Appendix~\ref{app:recipe}), and we report training-time hyperparameter
comparisons against optimizer-step counts rather than epochs so that the
comparison point is stated explicitly.

\paragraph{Single-step training with autoregressive evaluation shows
exposure bias in our runs.} Under teacher
forcing the same model stays stable over 20 steps, while under
autoregression its increments self-amplify by roughly a factor of seven.
Single-step accuracy and rollout stability are not tightly coupled:
two checkpoints differing by $0.1\%$ in 20-step mean nRMSE differed by a
factor of three in enstrophy ratio, and the decoupling runs in both
directions across seeds, with one run converging to the best validation
loss of its group ($3.6\times10^{-4}$) yet diverging under
autoregression to nRMSE $26.8$, while another stalled at a higher
validation plateau ($1.0\times10^{-3}$) yet rolled out healthily at
$0.79$, ahead of persistence. Validation loss therefore cannot select
rollout models. A single averaged error is likewise blind to
this, which is the direct motivation for the three-axis leaderboard and the
per-case skill score (Section~\ref{sec:benchmark}).

\paragraph{We audited the baselines that failed, not only those that
succeeded.} A failure to learn can be a reproduction defect rather than an
architectural verdict: one attention baseline initially failed to learn
on every configuration. The cause was a single missing normalization step relative
to the official implementation, which amplified the layer signal by a large
factor; once restored, the model competed normally. A failure to learn can
be a reproduction defect rather than an architectural verdict, so all five
baselines carry the deviation audit of Appendix~\ref{app:deviations}.

\paragraph{The metric layer is audited on real fields.}
Every metric and non-learning reference is targeted against physical
invariants of real DNS fields, not only synthetic fields, because the known-answer
synthetic fields we use do not reproduce the forward cascade or the
super-Gaussian tails that the metrics measure. A family of unit and
axis-order defects passed synthetic tests while producing meaningless
values on turbulence; corpus-level tests with mutation coverage now require
that reverting any such defect turns a real-data test red.

\section{Specified Tasks: Denoising and Parameter Inversion}
\label{app:planned-tasks}
P6 (denoising: $X+\eta\rightarrow X$ under synthetic measurement noise)
and P7 (parameter inversion: infer $(Re_\lambda,\nu)$ under fixed forcing) are
specified but not part of the reference results; their protocols are fixed
here so future results are comparable.

\section{Complete Benchmark Results}
\label{app:results}
This appendix lists the per-configuration result matrices behind
Section~\ref{sec:results}, extracted from the released per-run result
files; every number is reproducible from the released evaluation output
shipped with the benchmark. Additional per-run side metrics
(high-band spectral errors, vorticity-gradient statistics, SGS
energy-transfer rates) are included in the released result files but
not tabulated here.

\paragraph{High-band spectral error.}
Table~\ref{tab:app-highband} tabulates the third leaderboard axis per
configuration and model behind the aggregate contrast quoted in
Section~\ref{sec:three-axis}.

\begin{table}[H]
\centering
\caption{High-band spectral error at rollout step 20 (median over the training seeds of Table~\ref{tab:app-seeds}), the third
leaderboard axis: absolute RMS of Fourier-mode differences in the top
wavenumber band, computed after de-normalization so that it carries
physical units and stays comparable across configurations (lower is
better). Persistence, not tabulated, stays at $\le0.1$ on every
configuration: copying a real snapshot forward
keeps a physically correct high-band amplitude, whereas the learned
operators inject spurious high-wavenumber energy.}
\label{tab:app-highband}
\footnotesize
\setlength{\tabcolsep}{5pt}
\begin{tabular}{@{}lcccc@{}}
\toprule
\textbf{Config} & \textbf{FNO3d} & \textbf{TFNO} & \textbf{Transolver} & \textbf{Spectral U-Net} \\
\midrule
$k_f{=}3$       & 8.8  & 5250  & 442.0  & 124.9 \\
$k_f{=}4$       & 0.9  & 0.3   & 359.5  & 87.8  \\
$\tau{=}1$      & 25.3 & 481.6 & 596.3  & 144.1 \\
$Re_\lambda 55$ & 5.1  & 0.4   & 399.7  & 39.0  \\
$Re_\lambda 70$ & 14.5 & 2951  & 533.4  & 87.1  \\
$Re_\lambda 86$ & 47.9 & 311.3 & 1309.5 & 419.5 \\
\bottomrule
\end{tabular}
\end{table}

\paragraph{Rollout on the extended-physics and free-decay
configurations.}
Table~\ref{tab:app-rollout-ext} extends the forecasting leaderboard
beyond the main table.

\begin{table*}[t]\centering
\caption{Forecasting (P1), extended-physics and free-decay configurations
(seed 0): nRMSE (mean over the 20 rollout steps) / final-step enstrophy
ratio / EPT. $^{\dagger}$Divergence: 20-step mean nRMSE above 1 on the
forced rows; above $1.2\times$ persistence, or an enstrophy ratio above
100, on the decay rows, whose persistence error itself approaches 1.
Bold marks, per row and among the four learned models, the lowest nRMSE,
the enstrophy ratio closest to one, and the longest EPT; ties at the
printed precision are resolved at full precision, and exact ties are
left unbolded. Scored windows per configuration: 51 for the rotating
pair and helical, 37 for the passive scalar, 12 for hot-start decay and
ABC, 6 for the two cold-start decay families (window-limited).}
\label{tab:app-rollout-ext}\footnotesize\setlength{\tabcolsep}{3.2pt}
\begin{tabular}{@{}lccccc@{}}
\toprule
Config & FNO3d & TFNO & Transolver & Spectral U-Net & persistence \\
\midrule
rotating (strong) & \textbf{0.475} / \textbf{17.0} / \textbf{6.2} & 0.887 / 1721.8 / 4.3 & 0.488 / 27.1 / 1.5 & 0.497 / 44.5 / 2.5 & 0.491 \\
rotating (moderate) & 0.808 / \textbf{6.2} / 2.2 & 1.174$^{\dagger}$ / 110.6 / \textbf{2.3} & 0.700 / 25.8 / 0.4 & \textbf{0.695} / 94.0 / 0.9 & 0.734 \\
passive scalar & 0.886 / \textbf{1.9} / 0.8 & 1.586$^{\dagger}$ / 62.2 / \textbf{1.4} & \textbf{0.844} / 16.0 / 0.0 & 0.922 / 50.7 / 0.4 & 0.884 \\
helical & 0.837 / \textbf{4.1} / 1.9 & 7.227$^{\dagger}$ / 14357.9 / \textbf{2.0} & \textbf{0.781} / 16.3 / 0.4 & 0.839 / 43.1 / 0.8 & 0.818 \\
decay (hot-start) & \textbf{0.841} / 0.2 / 0.1 & 0.844 / \textbf{0.3} / 0.1 & 0.987$^{\dagger}$ / 117.0 / 0.0 & 1.023$^{\dagger}$ / 195.8 / \textbf{0.3} & 1.017 \\
decay (Saffman) & 0.887 / 2.1 / 0.7 & \textbf{0.880} / \textbf{2.0} / \textbf{0.7} & 1.235$^{\dagger}$ / 701.0 / 0.7 & 1.486$^{\dagger}$ / 385.0 / 0.4 & 0.899 \\
decay (Batchelor) & \textbf{0.786} / 2.0 / \textbf{1.4} & 0.787 / \textbf{2.0} / 1.4 & 1.343$^{\dagger}$ / 991.5 / 1.3 & 1.560$^{\dagger}$ / 306.6 / 0.6 & 0.796 \\
decay (ABC) & 0.069 / 0.98 / 19.0 & \textbf{0.069} / \textbf{0.99} / 19.0 & 0.075 / 35.0 / 19.0 & 0.205$^{\dagger}$ / 85.0 / 14.5 & 0.069 \\
\bottomrule
\end{tabular}
\end{table*}

Strong rotation is the easiest of the driven settings (persistence
$0.491$), where FNO3d and Transolver sit marginally below
persistence while TFNO and the Spectral U-Net stay above it, and the
quasi-steady ABC case leaves almost no headroom by design. The
free-decay rows invert the stability ordering of
Section~\ref{sec:rollout} twice over. Under the decay divergence rule
Transolver diverges on three of the four decay families (enstrophy
ratios $117$--$991$), and on hot-start on all three seeds; the Spectral
U-Net diverges on all four, including the quasi-steady ABC case where
its error is three times persistence's, despite its seed stability under
forcing (Section~\ref{sec:rollout}). TFNO reverses in the opposite
direction: seed-fragile under forcing, it is the steadiest model on
the three non-quasi-steady decay families ($0.79$--$0.88$, enstrophy
ratios between $0.3$ and $2.0$), where FNO3d matches it. Architectural stability is therefore
regime-dependent rather than a property of the operator family.

\paragraph{Super-resolution.}
Spectral interpolation is a strong trivial reference for
this task (Tables~\ref{tab:app-superres} and~\ref{tab:app-superres-ext}).
FNO3d improves on it across the main-table configurations by
$0.012$--$0.048$ in nRMSE, but on the extended-physics and free-decay
configurations six of the eight sit within $0.015$ of the reference in
either direction, and the reference is clearly ahead on the other two,
by a factor of two under strong rotation and by a third on the
quasi-steady ABC case.

\begin{table}[H]\centering
\caption{Super-resolution (P2), main-table configurations: nRMSE / spectrum-$L_2$. $^{\S}$Trained at frame stride 1, four times the training frames per epoch of the forecasting recipe.}
\label{tab:app-superres}\footnotesize\setlength{\tabcolsep}{2.2pt}
\begin{tabular}{@{}lccccc@{}}
\toprule
Config & FNO3d & TFNO & Transolver & Spectral U-Net & spectral interp. \\
\midrule
$k_f{=}3$ & 0.186 / 0.022$^{\S}$ & 0.215 / 0.028 & 0.213 / 0.056 & 0.287 / 0.209$^{\S}$ & 0.221 / 0.028 \\
$k_f{=}4$ & 0.198 / 0.027 & 0.222 / 0.058 & 0.214 / 0.045 & 0.306 / 0.197 & 0.246 / 0.042 \\
$\tau{=}1$ & 0.180 / 0.020 & 0.307 / 0.098 & 0.221 / 0.059 & 0.296 / 0.266 & 0.197 / 0.019 \\
$Re_\lambda 55$ & 0.133 / 0.011 & 0.194 / 0.021 & 0.176 / 0.036 & 0.254 / 0.283 & 0.145 / 0.012 \\
$Re_\lambda 70$ & 0.151 / 0.017 & 0.271 / 0.072 & 0.202 / 0.036 & 0.269 / 0.204 & 0.171 / 0.016 \\
$Re_\lambda 86$ & 0.178 / 0.017 & 0.269 / 0.064 & 0.224 / 0.051 & 0.286 / 0.229 & 0.197 / 0.018 \\
\bottomrule
\end{tabular}
\end{table}

Two structures sit beneath that summary. First, the ordering
among the learned models is fixed: FNO3d has the lowest nRMSE on every
one of the fourteen configurations carrying the task, in contrast to
forecasting, where the leader changes with the configuration; the axes
still disagree, though, since on the cold-start families FNO3d leads
nRMSE while TFNO leads spectrum-$L_2$
(Table~\ref{tab:app-superres-ext}). Second, the Spectral U-Net collapses on the cold-start decays,
reaching nRMSE $0.510$ and $0.552$ where the other learned models stay
at $0.25$--$0.31$, with spectrum-$L_2$ an order of magnitude above
FNO3d's ($0.76$ and $0.86$ against $0.08$ and $0.11$); on the driven
configurations it is merely the weakest of the three
($0.30$--$0.35$). As with the pressure and subgrid-stress failures of
Section~\ref{sec:results}, the failure is architecture- and
regime-specific rather than a property of the task. The spectrum-$L_2$
column favors the learned models less than nRMSE does.

\begin{table}[H]\centering
\caption{Super-resolution (P2), extended-physics and free-decay
configurations: nRMSE / spectrum-$L_2$. Bold marks, per row and among
the three learned models, the lowest value of each metric; spectral
interpolation is a trivial reference and is not ranked.}
\label{tab:app-superres-ext}\footnotesize\setlength{\tabcolsep}{3.2pt}
\begin{tabular}{@{}lcccc@{}}
\toprule
Config & FNO3d & TFNO & Spectral U-Net & spectral interp. \\
\midrule
rotating (strong) & \textbf{0.155} / \textbf{0.063} & 0.245 / 0.065 & 0.345 / 0.194 & 0.070 / 0.002 \\
rotating (moderate) & \textbf{0.129} / \textbf{0.020} & 0.207 / 0.042 & 0.297 / 0.217 & 0.133 / 0.007 \\
passive scalar & \textbf{0.219} / \textbf{0.023} & 0.272 / 0.060 & 0.315 / 0.238 & 0.222 / 0.019 \\
helical & \textbf{0.186} / \textbf{0.022} & 0.229 / 0.028 & 0.295 / 0.301 & 0.187 / 0.016 \\
decay (hot-start) & \textbf{0.224} / \textbf{0.035} & 0.311 / 0.119 & 0.366 / 0.378 & 0.236 / 0.034 \\
decay (Saffman) & \textbf{0.250} / 0.079 & 0.263 / \textbf{0.049} & 0.510 / 0.759 & 0.235 / 0.047 \\
decay (Batchelor) & \textbf{0.288} / 0.106 & 0.308 / \textbf{0.079} & 0.552 / 0.855 & 0.283 / 0.071 \\
decay (ABC) & \textbf{0.065} / \textbf{0.004} & 0.090 / 0.007 & 0.265 / 0.090 & 0.043 / 0.002 \\
\bottomrule
\end{tabular}
\end{table}

\paragraph{Sparse reconstruction.}
Reconstruction from 5\% of observed points is the task the learned
models win most clearly (Tables~\ref{tab:app-recon}
and~\ref{tab:app-recon-ext}): against the do-nothing identity reference of
$0.975$, Transolver and the Spectral U-Net reconstruct to
$0.21$--$0.27$ on the main-table configurations, FNO3d to
$0.27$--$0.33$, while TFNO is erratic across configurations
($0.32$--$0.79$). Transolver carries that ability across every extended
and decay configuration ($0.06$--$0.37$), and TFNO is again the weakest
wherever it runs ($0.38$--$0.80$).

\begin{table}[H]\centering
\caption{Sparse reconstruction (P3), main-table configurations: nRMSE. $^{\S}$Trained at frame stride 1, four times the training frames per epoch of the forecasting recipe.}
\label{tab:app-recon}\footnotesize\setlength{\tabcolsep}{4pt}
\begin{tabular}{@{}lccccc@{}}
\toprule
Config & FNO3d & TFNO & Transolver & Spectral U-Net & identity \\
\midrule
$k_f{=}3$ & 0.327 & 0.446 & 0.259$^{\S}$ & 0.216$^{\S}$ & 0.975 \\
$k_f{=}4$ & 0.322 & 0.324 & 0.272 & 0.214 & 0.975 \\
$\tau{=}1$ & 0.316 & 0.783 & 0.250 & 0.237 & 0.975 \\
$Re_\lambda 55$ & 0.269 & 0.352 & 0.211 & 0.218 & 0.975 \\
$Re_\lambda 70$ & 0.281 & 0.785 & 0.231 & 0.218 & 0.975 \\
$Re_\lambda 86$ & 0.317 & 0.535 & 0.250 & 0.225 & 0.975 \\
\bottomrule
\end{tabular}
\end{table}

\begin{table}[H]\centering
\caption{Sparse reconstruction (P3), extended-physics and free-decay
configurations: nRMSE (seed 0; TFNO and Spectral U-Net on 60 samples,
Transolver on 51). $^{\ast}$Batchelor test
trajectories contain one jump in physical time; the Spectral U-Net value mixes
0.265 before it and 0.742 after.}
\label{tab:app-recon-ext}\footnotesize\setlength{\tabcolsep}{4pt}
\begin{tabular}{@{}lcccc@{}}
\toprule
Config & TFNO & Transolver & Spectral U-Net & identity \\
\midrule
rotating (strong) & 0.376 & 0.133 & 0.289 & 0.975 \\
rotating (moderate) & 0.550 & 0.208 & 0.263 & 0.975 \\
passive scalar & --- & 0.283 & --- & 0.975 \\
helical & --- & 0.250 & 0.244 & 0.975 \\
decay (hot-start) & 0.802 & 0.278 & 0.316 & 0.975 \\
decay (Saffman) & 0.662 & 0.314 & 0.440 & 0.975 \\
decay (Batchelor) & --- & 0.372 & 0.631$^{\ast}$ & 0.975 \\
decay (ABC) & --- & 0.059 & 0.255 & 0.975 \\
\bottomrule
\end{tabular}
\end{table}

\paragraph{Pressure recovery.}
The spectral Poisson solve is exact by construction and no learned
model approaches it (Tables~\ref{tab:app-pressure}
and~\ref{tab:app-pressure-ext}): FNO3d, the best learned model, spans
$0.45$--$0.87$ across the forced configurations and the hot-start decay,
reaches the zero-field level ($\approx\!1.0$) on the cold-start decays,
and DeepONet-3D
sits near that level on every configuration except the quasi-steady
ABC case ($0.66$), for the encoding reasons given in
Section~\ref{sec:results}.

\begin{table}[H]\centering
\caption{Pressure recovery (P4), main-table configurations: nRMSE / Poisson residual. $^{\S}$Trained at frame stride 1, four times the training frames per epoch of the forecasting recipe.}
\label{tab:app-pressure}\footnotesize\setlength{\tabcolsep}{4pt}
\begin{tabular}{@{}lcccc@{}}
\toprule
Config & FNO3d & TFNO & DeepONet-3D & Poisson solve \\
\midrule
$k_f{=}3$ & 0.467 / 1.85$^{\S}$ & 0.966 / 8.34 & 1.003 / 1.12$^{\S}$ & 0.000 / 0.02 \\
$k_f{=}4$ & 0.446 / 1.70 & 0.962 / 6.34 & 1.000 / 1.06 & 0.000 / 0.01 \\
$\tau{=}1$ & 0.668 / 1.43 & 0.965 / 25.07 & 1.005 / 1.29 & 0.000 / 0.02 \\
$Re_\lambda 55$ & 0.635 / 2.63 & 0.953 / 45.22 & 1.001 / 1.52 & 0.000 / 0.00 \\
$Re_\lambda 70$ & 0.719 / 1.33 & 0.973 / 32.97 & 1.001 / 1.37 & 0.000 / 0.00 \\
$Re_\lambda 86$ & 0.791 / 1.18 & 0.967 / 22.68 & 0.999 / 1.10 & 0.000 / 0.02 \\
\bottomrule
\end{tabular}
\end{table}

Errors are an order of magnitude lower on
the rotating configurations ($0.03$--$0.19$,
Table~\ref{tab:app-pressure-ext}), where large-scale
rotational balance dominates the pressure field; TFNO approaches FNO3d
there in nRMSE with a Poisson residual 19 to 29 times larger, another
case of pointwise error and physical consistency coming apart.

\begin{table}[H]\centering
\caption{Pressure recovery (P4), extended-physics and free-decay
configurations: nRMSE / Poisson residual (seed 0). FNO3d entries use 51
samples, TFNO and DeepONet-3D entries 60.}
\label{tab:app-pressure-ext}\footnotesize\setlength{\tabcolsep}{2.4pt}
\resizebox{\linewidth}{!}{\begin{tabular}{@{}lcccc@{}}
\toprule
Config & FNO3d & TFNO & DeepONet-3D & Poisson solve \\
\midrule
rotating (strong) & 0.029 / 14.23 & 0.051 / 269.9 & 0.993 / 34.5 & 0.000 / 0.00 \\
rotating (moderate) & 0.186 / 2.01 & 0.192 / 57.6 & 1.087 / 8.18 & 0.000 / 0.02 \\
passive scalar & 0.729 / 1.23 & --- & --- & 0.000 / 0.03 \\
helical & 0.799 / 1.23 & --- & 1.010 / 1.80 & 0.000 / 0.02 \\
decay (hot-start) & 0.870 / 1.41 & 0.980 / 16.18 & 1.010 / 1.39 & 0.000 / 0.02 \\
decay (Saffman) & 1.000 / 1.00 & 1.042 / 56.5 & 1.017 / 1.60 & 0.000 / 0.01 \\
decay (Batchelor) & 1.001 / 1.00 & --- & 1.028 / 1.72 & 0.000 / 0.01 \\
decay (ABC) & 0.098 / 13.78 & --- & 0.659 / 286.1 & 0.000 / 0.00 \\
\bottomrule
\end{tabular}}
\end{table}

\paragraph{Subgrid stress.}
FNO3d and TFNO roughly triple dynamic Smagorinsky's stress
correlation ($0.45$--$0.54$ against $0.16$--$0.17$) and report the
non-zero backscatter fraction that an eddy-viscosity closure cannot
express by construction (Tables~\ref{tab:app-sgs}
and~\ref{tab:app-sgs-ext}).

\begin{table*}[t]\centering
\caption{Subgrid-stress closure (P5), main-table configurations: nRMSE / stress correlation / backscatter fraction. Dynamic Smagorinsky predicts the trace-free stress and is scored against the full stress, which lowers its correlation by about $0.09$. $^{\S}$Trained at frame stride 1, four times the training frames per epoch of the forecasting recipe.}
\label{tab:app-sgs}\footnotesize\setlength{\tabcolsep}{4pt}
\begin{tabular}{@{}lcccc@{}}
\toprule
Config & FNO3d & TFNO & Spectral U-Net & dyn.\ Smagorinsky \\
\midrule
$k_f{=}3$ & 0.767 / 0.51 / 0.44$^{\S}$ & 0.764 / 0.52 / 0.51 & 1.017 / 0.23 / 0.45$^{\S}$ & 0.996 / 0.17 / 0.00 \\
$k_f{=}4$ & 0.747 / 0.53 / 0.46 & 0.746 / 0.54 / 0.53 & 0.931 / 0.28 / 0.44 & 0.995 / 0.17 / 0.00 \\
$\tau{=}1$ & 0.799 / 0.49 / 0.44 & 0.797 / 0.49 / 0.51 & 1.119 / 0.24 / 0.46 & 0.996 / 0.16 / 0.00 \\
$Re_\lambda 55$ & 0.823 / 0.45 / 0.50 & 0.822 / 0.45 / 0.50 & 1.262 / 0.15 / 0.47 & 0.992 / 0.17 / 0.00 \\
$Re_\lambda 70$ & 0.807 / 0.47 / 0.48 & 0.804 / 0.47 / 0.49 & 1.306 / 0.18 / 0.46 & 0.994 / 0.17 / 0.00 \\
$Re_\lambda 86$ & 0.783 / 0.50 / 0.47 & 0.782 / 0.50 / 0.50 & 1.186 / 0.22 / 0.47 & 0.996 / 0.17 / 0.00 \\
\bottomrule
\end{tabular}
\end{table*}

The pattern holds across every
extended-physics configuration, with the quasi-steady ABC case
reaching a correlation of $0.82$ for FNO3d
(Table~\ref{tab:app-sgs-ext}). TFNO stays below nRMSE 1 on all seven
extended configurations where it runs ($0.61$--$0.84$), and below the
same-protocol dynamic Smagorinsky on hot-start, Saffman, and Batchelor,
while the Spectral U-Net stays above 1 on the same seven
($1.26$--$11.0$) with the lowest stress correlations; the outcome
separates by model rather than by configuration. The released files also carry the net
resolved-to-subgrid energy-transfer rate per cell, spanning
$-3.4\times10^{-4}$ to $0.10$ across the reference matrix.

\begin{table*}[t]\centering
\caption{Subgrid-stress closure (P5), extended-physics and free-decay
configurations: nRMSE / stress correlation / backscatter fraction
(seed 0). FNO3d entries use 51 samples; TFNO, Spectral U-Net, and the
Smagorinsky entries of the hot-start, Saffman, and Batchelor rows use
60. n/d: not determined, because the
FNO3d decay evaluation applied the velocity energy floor to the stress
output; its correlation and backscatter are unaffected. Dynamic Smagorinsky is
scored against the full stress, as in Table~\ref{tab:app-sgs}.}
\label{tab:app-sgs-ext}\footnotesize\setlength{\tabcolsep}{5pt}
\begin{tabular}{@{}lcccc@{}}
\toprule
Config & FNO3d & TFNO & Spectral U-Net & dyn.\ Smagorinsky \\
\midrule
rotating (strong) & 0.837 / 0.46 / 0.58 & 0.832 / 0.45 / 0.52 & 2.061 / 0.18 / 0.50 & 0.997 / 0.09 / 0.00 \\
rotating (moderate) & 0.812 / 0.47 / 0.48 & 0.812 / 0.46 / 0.52 & 1.518 / 0.16 / 0.49 & 0.993 / 0.15 / 0.00 \\
passive scalar & 0.777 / 0.50 / 0.44 & --- & --- & 0.996 / 0.17 / 0.00 \\
helical & 0.790 / 0.49 / 0.45 & 0.787 / 0.49 / 0.51 & 1.405 / 0.14 / 0.47 & 0.995 / 0.17 / 0.00 \\
decay (hot-start) & n/d / 0.51 / 0.49 & 0.788 / 0.52 / 0.53 & 1.264 / 0.21 / 0.46 & 0.997 / 0.17 / 0.00 \\
decay (Saffman) & n/d / 0.55 / 0.52 & 0.782 / 0.55 / 0.50 & 2.757 / 0.08 / 0.48 & 0.996 / 0.17 / 0.00 \\
decay (Batchelor) & n/d / 0.59 / 0.49 & 0.841 / 0.59 / 0.50 & 3.010 / 0.11 / 0.48 & 0.996 / 0.17 / 0.00 \\
decay (ABC) & 0.476 / 0.82 / 0.47 & 0.612 / 0.69 / 0.53 & 11.00 / 0.22 / 0.48 & 0.999 / 0.04 / 0.00 \\
\bottomrule
\end{tabular}
\end{table*}

\begin{figure}[t]
  \centering
  \includegraphics[width=\linewidth]{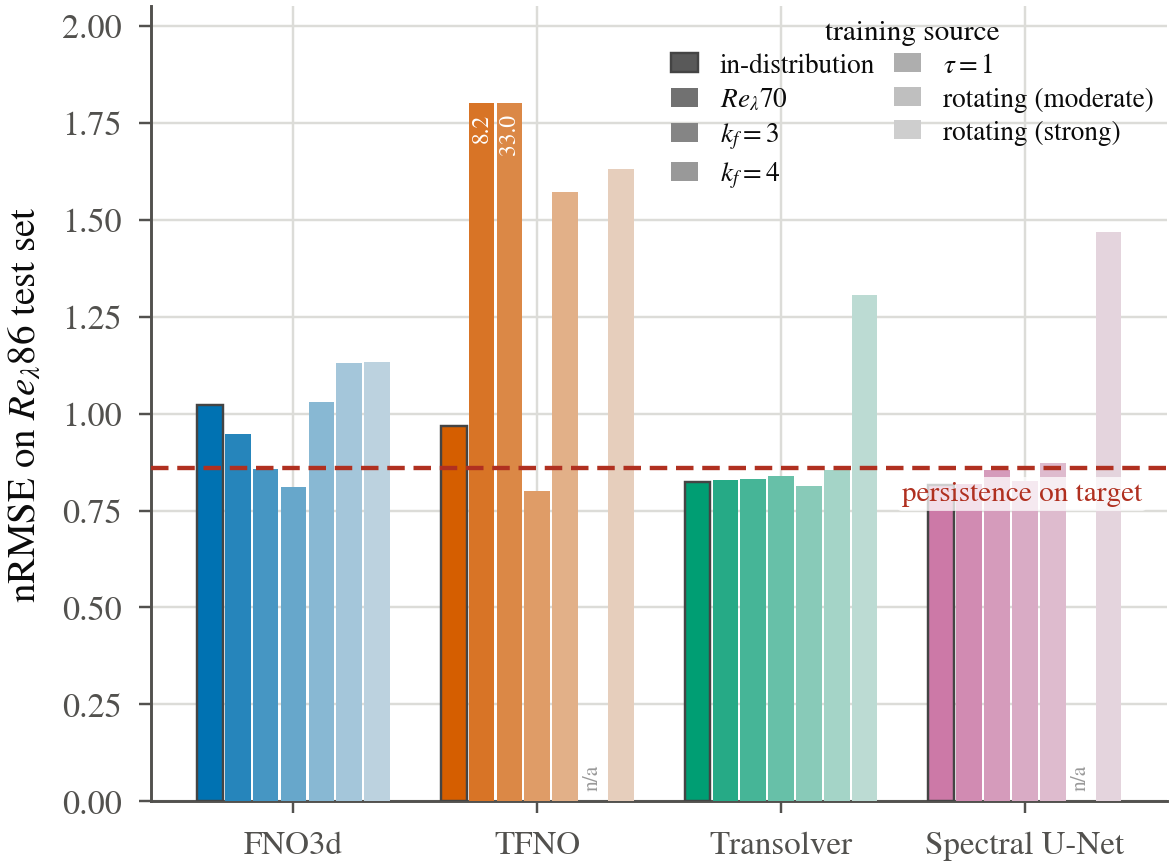}
  \caption{Zero-shot transfer onto the $Re_\lambda 86$ test set (all 135
  windows, target normalization), sources ordered by physical distance;
  bars cut at 1.8 carry their value.}
  \label{fig:ggroup-bars}
\end{figure}

\paragraph{Generalization transfers.}
The transfer grid onto the flagship test set (four-model transfers from
the Reynolds and forcing neighbors, and the two-model
rotation-specialized and cross-physics transfers) is reported in Table~\ref{tab:app-ood} (main text) and
Table~\ref{tab:app-ood-ext}; Figure~\ref{fig:ggroup-bars} orders the
flagship transfers by physical distance. The two
further directions of Section~\ref{sec:results} are tabulated below.

\begin{table}[H]\centering
\caption{Rotation-specialized and cross-physics transfers: 20-step
nRMSE, target / source normalization. Rotating sources are scored on all
135 $Re_\lambda 86$ windows; the scalar$\rightarrow$stratified probe uses
five-channel models on 51 windows under the target normalization.}
\label{tab:app-ood-ext}\footnotesize\setlength{\tabcolsep}{2.6pt}
\begin{tabular}{@{}lcccc@{}}
\toprule
Source & FNO3d & TFNO & Transolver & Spectral U-Net \\
\midrule
rotating (strong) & 1.132 / 1.131 & 1.631 / 1.858 & 1.305 / 1.108 & 1.468 / 1.353 \\
rotating (moderate) & 1.131 / 1.068 & --- & 0.854 / 0.856 & --- \\
in-distribution & 1.022 & 0.969 & 0.823 & 0.816 \\
persistence & 0.860 & 0.860 & 0.860 & 0.860 \\
\midrule
scalar$\to$stratified & 0.944 & --- & 0.854 & --- \\
persistence (stratified) & 0.909 & 0.909 & 0.909 & 0.909 \\
\bottomrule
\end{tabular}
\end{table}

The first is transfer along the forcing axis onto the $k_f{=}4$ test set
(Table~\ref{tab:app-g3}), where FNO3d and Transolver arrive from the
neighboring band within $0.04$ of, or better than, their
in-distribution error, the Spectral U-Net $0.08$ above it, and TFNO
carries over the divergence of its source.

\begin{table}[H]\centering
\caption{Zero-shot transfer along the forcing axis onto the $k_f{=}4$
test set (20-step nRMSE, 51 windows; target / source normalization where
both were scored). Seed-0 sources under the reference recipe.}
\label{tab:app-g3}\footnotesize\setlength{\tabcolsep}{3pt}
\begin{tabular}{@{}lcccc@{}}
\toprule
Source & FNO3d & TFNO & Transolver & Spectral U-Net \\
\midrule
$k_f{=}3$ & 0.700 / 0.649 & 23.4 / 13.3 & 0.789 / 0.789 & 0.920 / 1.008 \\
$Re_\lambda 86$ ($k_f{=}2$) & 0.883 & 0.924 & 0.787 & 0.905 \\
\midrule
in-distribution & 0.661 & 0.759 & 0.800 & 0.840 \\
persistence & 0.822 & 0.822 & 0.822 & 0.822 \\
\bottomrule
\end{tabular}
\end{table}

The second is transfer from the generic forced flows
onto the rotating test sets (Table~\ref{tab:app-g5rev}), the reverse of
the specialization direction, where FNO3d degrades in both
directions, Transolver transfers without loss under both
normalizations, and the Spectral U-Net does so under the source model's
normalization ($0.430$ against its in-distribution $0.497$) while
degrading mildly under the target's ($0.634$--$0.685$). The
in-distribution row qualifies one column: TFNO fails on strong rotation
in distribution as well ($0.887$ with a diverged enstrophy ratio), so its
transfer entries there measure a regime the model does not learn rather
than a transfer loss.

\begin{table}[H]\centering
\caption{Zero-shot transfer (20-step nRMSE, 51 windows) from the generic
forced flows onto the rotating test sets, the reverse of the
specialization direction in Table~\ref{tab:app-ood-ext}; target / source
normalization where both were scored.}
\label{tab:app-g5rev}\footnotesize\setlength{\tabcolsep}{2.6pt}
\begin{tabular}{@{}lcccc@{}}
\toprule
Source $\to$ target & FNO3d & TFNO & Transolver & Spectral U-Net \\
\midrule
$k_f{=}3\to$ rot.\ (strong) & 1.081 / 0.833 & 10.8 / 27.7 & 0.485 / 0.323 & 0.685 / 0.430 \\
$Re_\lambda 86\to$ rot.\ (strong) & 1.500 & 1.715 & 0.501 & 0.634 \\
$Re_\lambda 86\to$ rot.\ (moderate) & 1.238 & 1.126 & 0.720 & 0.836 \\
\midrule
in-distribution (strong) & 0.475 & 0.887 & 0.488 & 0.497 \\
persistence (strong) & 0.491 & 0.491 & 0.491 & 0.491 \\
persistence (moderate) & 0.734 & 0.734 & 0.734 & 0.734 \\
\bottomrule
\end{tabular}
\end{table}

The forced-to-decay axis, which is not scored as a transfer, is
measured in Appendix~\ref{app:conditioning}.

\section{Extended Results}
\label{app:extended-results}
\begin{figure}[t]
  \centering
  \includegraphics[width=\linewidth]{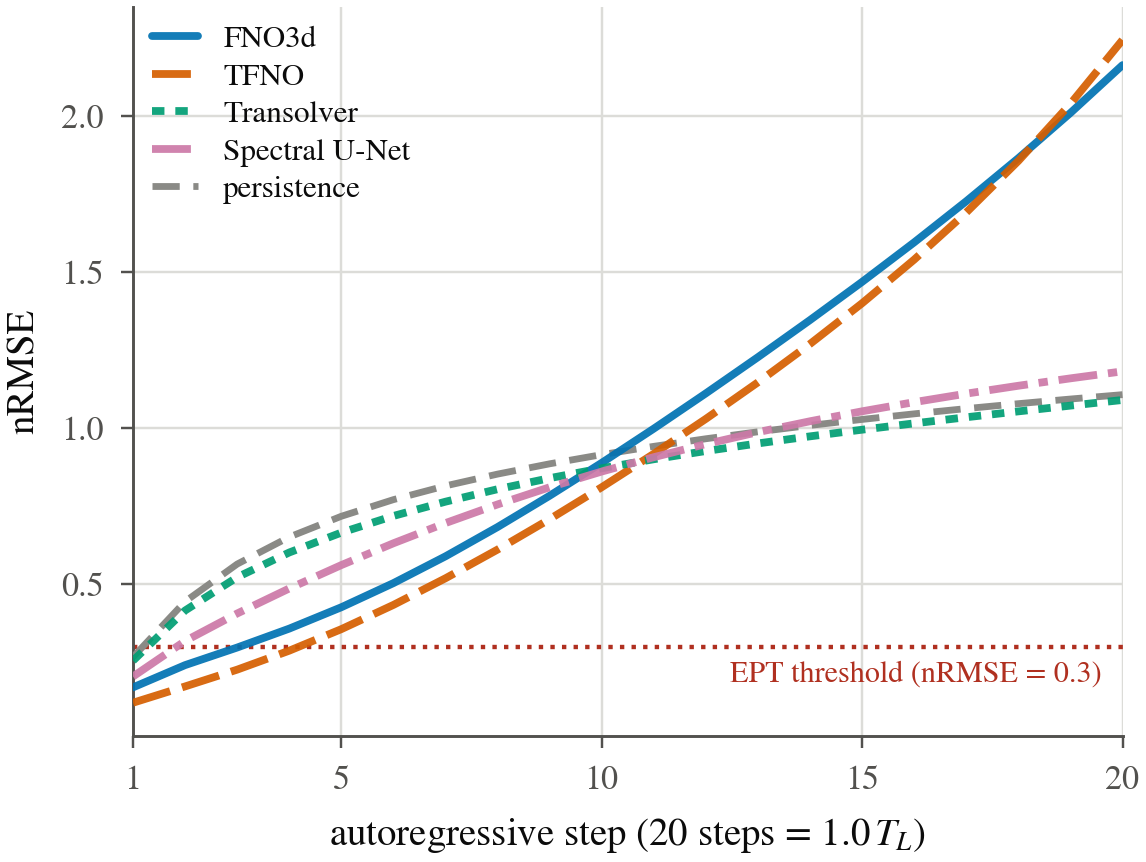}
  \caption{Per-step rollout error on $Re_\lambda 86$ (seed 0, all 135
  test windows); EPT = crossing of nRMSE $0.3$ (one autoregressive step =
  one frame = $0.05\,T_L$).}
  \label{fig:rollout-curves}
\end{figure}

\paragraph{Rollout stability, per configuration.}
Figure~\ref{fig:rollout-curves} traces per-step error over the 20-step
rollout on the flagship. TFNO and FNO3d start lowest (first-step error
$0.45$ and $0.64$ of persistence's) and cross persistence near step 10,
while Transolver starts close to persistence and flattens. On the
free-decay configurations the ordering reverses, with Transolver's
enstrophy ratio reaching $991$ while FNO3d stays bounded.

\begin{table*}[t]
\centering
\caption{Forecasting (P1) across training seeds: 20-step nRMSE as median
[min--max] (diverged / seeds), reference recipe, 51 windows per
configuration. The last row scores the flagship on all 135 windows of
its seven test trajectories (four seeds).}
\label{tab:app-seeds}
\footnotesize
\setlength{\tabcolsep}{4pt}
\begin{tabular}{@{}lcccc@{}}
\toprule
\textbf{Config} & \textbf{FNO3d} & \textbf{TFNO} & \textbf{Transolver} & \textbf{Spectral U-Net} \\
\midrule
$k_f{=}3$       & 0.779 [0.771--0.810] (0/5) & 14.0 [3.27--2582] (5/5) & 0.810 [0.810--0.811] (0/5) & 0.830 [0.819--0.843] (0/5) \\
$k_f{=}4$       & 0.672 [0.661--0.709] (0/5) & 0.763 [0.749--0.765] (0/5) & 0.799 [0.799--0.800] (0/5) & 0.851 [0.831--0.863] (0/5) \\
$\tau{=}1$      & 1.149 [0.992--1.224] (4/5) & 1.838 [1.592--3.71] (5/5) & 0.843 [0.841--0.844] (0/3) & 0.860 [0.860--0.862] (0/3) \\
$Re_\lambda 55$ & 0.905 [0.761--1.117] (1/5) & 0.786 [0.781--8.72] (1/5) & 0.803 [0.803--0.806] (0/5) & 0.767 [0.754--0.772] (0/5) \\
$Re_\lambda 70$ & 0.929 [0.798--1.091] (2/5) & 9.97 [0.867--26.8] (4/5) & 0.789 [0.788--0.789] (0/5) & 0.804 [0.792--0.817] (0/5) \\
$Re_\lambda 86$ & 1.065 [0.984--1.155] (4/5) & 1.371 [0.940--3.51] (3/5) & 0.819 [0.816--0.821] (0/5) & 0.820 [0.813--0.830] (0/5) \\
\midrule
$Re_\lambda 86$, 135 windows & 1.049 [0.994--1.097] (3/4) & 1.235 [0.969--4.02] (2/4) & 0.826 [0.823--0.828] (0/4) & 0.816 [0.809--0.831] (0/4) \\
\bottomrule
\end{tabular}
\end{table*}

\begin{table*}[t]
\centering
\caption{Forecasting (P1) with denser training sampling: seed-0 models
trained on every frame (frame stride 1, four times the training pairs
per epoch of the reference recipe), otherwise identical. Cells: 20-step
nRMSE / EPT on the 51 reference windows; in brackets, the first-step
error relative to persistence at stride 4 $\rightarrow$ stride 1, for
the cells where both were measured on the same windows ($Re_\lambda 86$:
all 135 windows). $^{\dagger}$20-step mean nRMSE above 1.}
\label{tab:app-stride1}
\footnotesize
\setlength{\tabcolsep}{4pt}
\begin{tabular}{@{}lcccc@{}}
\toprule
\textbf{Config} & \textbf{FNO3d} & \textbf{TFNO} & \textbf{Transolver} & \textbf{Spectral U-Net} \\
\midrule
$k_f{=}3$       & 0.810 / 2.0 & 0.751 / 4.0 & 0.742 / 1.4 & 0.799 / 1.5 \\
$k_f{=}4$       & 0.663 / 2.3 [0.740$\rightarrow$0.735] & 1.587$^{\dagger}$ / 3.6 [0.995$\rightarrow$0.557] & 0.741 / 1.7 [0.999$\rightarrow$0.727] & 0.802 / 1.7 [0.874$\rightarrow$0.711] \\
$\tau{=}1$      & 1.254$^{\dagger}$ / 1.7 [0.662$\rightarrow$0.642] & 0.777 / 3.4 [0.520$\rightarrow$0.412] & 0.757 / 1.1 [0.993$\rightarrow$0.653] & 0.826 / 0.9 [0.827$\rightarrow$0.726] \\
$Re_\lambda 55$ & 0.846 / 3.1 & 1.152$^{\dagger}$ / 3.9 [0.917$\rightarrow$0.338] & 0.744 / 1.7 & 0.745 / 1.6 [0.841$\rightarrow$0.772] \\
$Re_\lambda 70$ & 0.877 / 2.9 & 0.765 / 4.5 & 0.719 / 1.6 & 0.772 / 1.3 \\
$Re_\lambda 86$ & 1.724$^{\dagger}$ / 2.3 [0.636$\rightarrow$0.620] & 0.711 / 4.5 [0.451$\rightarrow$0.383] & 0.706 / 1.6 [0.957$\rightarrow$0.536] & 0.752 / 1.4 [0.766$\rightarrow$0.624] \\
\bottomrule
\end{tabular}
\end{table*}

\paragraph{Seed variance.}
Table~\ref{tab:app-seeds} gives the seed ranges behind the medians of
Table~\ref{tab:main-results}, and Table~\ref{tab:app-stride1} the seed-0
models trained at frame stride 1. TFNO diverges on five of five seeds
on $k_f{=}3$ and $\tau{=}1$, four on $Re_\lambda 70$, and three on the
flagship, while it never diverges on $k_f{=}4$, where its first step
stays at the persistence level. Under full-trajectory scoring the order
of the two best models on $Re_\lambda 86$ reverses (Spectral U-Net $-$
Transolver $= -0.0076$, 95\% CI [$-0.0211$, $+0.0060$] with trajectories
resampled, lower on 4/7 trajectories; on a second seed the gap is
$-0.016$ [$-0.029$, $-0.002$], lower on 5/7), so the two models are
separable on one seed and not on the other. Scoring all 135 flagship
windows instead of the first 51 changes the four reference-seed models
by $-0.6\%$ to $+3.9\%$.
Of the 26 model pairs that share at least three seeds, 21 keep the same
sign on every seed and 16 are separable on every seed (10 and 5 of the
12 pairs without TFNO). On $\tau{=}1$ Transolver and the Spectral U-Net
are not separable (Transolver $-$ Spectral U-Net $= -0.016$ [$-0.036$,
$+0.005$] and $-0.017$ [$-0.036$, $+0.001$] on the two seeds they share).

\begin{table}[t]
\centering
\caption{Forecasting (P1) across three training seeds on three extended
configurations: 20-step nRMSE as median [min--max] (diverged / seeds),
reference recipe. Hot-start divergence follows the decay rule of
Table~\ref{tab:app-rollout-ext}; TFNO on helical is trained on one seed.}
\label{tab:app-seeds-ext}
\footnotesize
\setlength{\tabcolsep}{2pt}
\resizebox{\linewidth}{!}{\begin{tabular}{@{}lccc@{}}
\toprule
Model & rotating (strong) & helical & decay (hot-start) \\
\midrule
FNO3d & 0.475 [0.436--0.492] (0/3) & 0.837 [0.832--0.863] (0/3) & 0.893 [0.841--0.921] (0/3) \\
TFNO & 0.780 [0.623--0.887] (0/3) & 7.227 (1/1) & 0.844 [0.835--0.899] (0/3) \\
Transolver & 0.488 [0.487--0.492] (0/3) & 0.781 [0.780--0.782] (0/3) & 0.983 [0.978--0.987] (3/3) \\
Spectral U-Net & 0.497 [0.496--0.524] (0/3) & 0.839 [0.829--0.840] (0/3) & 1.023 [1.006--1.057] (1/3) \\
\midrule
persistence & 0.491 & 0.818 & 1.017 \\
\bottomrule
\end{tabular}}
\end{table}

\paragraph{Seeds on extended configurations.}
Table~\ref{tab:app-seeds-ext} repeats three configurations outside the
main table with three seeds. On strong rotation no model is below
persistence on every seed: FNO3d and Transolver are below it on two of
three (Transolver within $0.005$ of it), and TFNO and the Spectral U-Net
on none. On helical Transolver has the lowest error on every seed, while
FNO3d and the Spectral U-Net are above persistence on all three. On
hot-start FNO3d and TFNO stay below Transolver, which stays near
persistence, and the Spectral U-Net is highest on all three seeds, while
FNO3d and TFNO are not ordered; under the decay
rule Transolver diverges on all three through its enstrophy ratio
($105$--$120$) and the Spectral U-Net on one.

\paragraph{Where the spectral distortion sits.}
On the flagship configuration of Figure~\ref{fig:rollout-spectra} the
models fail in different bands. Transolver holds the mid-band spectrum
close to truth (energy ratio $0.78$ over $k\in[8,32]$) and concentrates
its excess in the far tail ($99\times$ truth over $k\in[48,85]$), while
FNO3d and TFNO carry a broadband excess (mid-band ratios $12$ and $17$),
with FNO3d's tail an order of magnitude closer to truth than
Transolver's and TFNO's between the two. The high-band spectral errors of Table~\ref{tab:app-highband} carry the same distinction across
configurations.

\paragraph{The trivial-to-informed interval.}
Section~\ref{sec:results} reports the bracket in aggregate; per
configuration the equations-informed integrator reaches nRMSE $0.410$,
$0.403$, $0.488$, $0.411$, $0.405$, and $0.441$ on $k_f{=}3$, $k_f{=}4$,
$\tau{=}1$, $Re_\lambda 55$, $Re_\lambda 70$, and $Re_\lambda 86$
respectively, against persistence values of $0.841$, $0.822$, $0.890$,
$0.831$, $0.822$, and $0.853$. On the hot-start decay, where no forcing
realization is hidden from it, the same integrator is near-exact (nRMSE
below $10^{-4}$), and Table~\ref{tab:app-integrator} splits its
forced-configuration error into the part an estimate of the current
forcing state removes and an upper bound on the rest.

\paragraph{Remaining open cells.}
On free decay, kinetic energy leaves the field, and Transolver and the
Spectral U-Net diverge there under the decay rule. On pressure, every network we benchmarked has higher error than
the exact spectral operator. These are stated as open challenges: cells
where no baseline in this suite exceeds its reference are where we see
headroom.

\paragraph{Efficiency.}
At matched capacity ($28$--$30.3$\,M parameters,
Table~\ref{tab:baselines}), the median run takes $0.2$--$0.9$\,h on a single
RTX~5090 (forecasting runs, the longest, $1.25$--$2.25$\,h), with peak
training memory between $2.8$ and $24.5$\,GB for the four models
reporting a peak, TFNO's being set by its checkpointing depth
(Table~\ref{tab:cost}). Transolver has both the highest mean
skill and the lowest wall-clock and memory in this suite, though the skill
margin over the Spectral U-Net (six-configuration mean of the seed
medians, $+0.038$ vs.\ $+0.025$) is smaller than most seed spreads; the Spectral U-Net has the highest
per-run cost.

\section{Protocol Studies}
\label{app:results-ablation}
The reference recipe of Section~\ref{sec:benchmark} fixes one choice per
protocol dimension. The studies below vary those choices one at a time on
representative configurations, all else fixed; each uses the reference
seed unless stated otherwise, and paired differences carry 95\% bootstrap intervals over
trajectories (within-trajectory blocks when fewer than seven) with the
number of trajectories on which the variant is lower.

\subsection{Training recipe}
\label{app:recipe}
\begin{table}[H]\centering
\caption{Learning rate on $k_f{=}3$ (seed 0, reference recipe otherwise;
51 test windows): 20-step nRMSE (final-step enstrophy ratio), and the
rate selected by 20-step rollout on the validation trajectory.}
\label{tab:app-lr}\footnotesize\setlength{\tabcolsep}{2.6pt}
\resizebox{\linewidth}{!}{\begin{tabular}{@{}lcccc@{}}
\toprule
Model & $10^{-3}$ (reference) & $3{\times}10^{-4}$ & $10^{-4}$ & validation \\
\midrule
FNO3d & 0.810 (4.53) & 0.752 (0.54) & 0.833 (0.92) & $3{\times}10^{-4}$ \\
TFNO & 42.3$^{\dagger}$ ($2.1{\times}10^{6}$) & 0.771 (0.41) & 0.831 (0.90) & $3{\times}10^{-4}$ \\
Transolver & 0.810 (8.8) & 0.864 (67.6) & 1.080$^{\dagger}$ (625) & $10^{-3}$ \\
\bottomrule
\end{tabular}}
\end{table}

\paragraph{Learning rate and validation selection.}
On $k_f{=}3$ (Table~\ref{tab:app-lr}) FNO3d's nRMSE varies by $0.08$
across the three rates while its enstrophy ratio ranges from $4.53$ to
$0.54$, and both lower rates fall outside the range of its five
reference seeds ($2.4$--$4.5$): the sensitivity is mild in nRMSE only.
FNO3d and TFNO also trade enstrophy against EPT, FNO3d's falling from
$1.98$ to $0.50$--$0.57$ frames at the lower rates, while Transolver and
the Spectral U-Net show no such trade. The model ordering depends on the
rate: FNO3d and Transolver differ by $0.0001$ at $10^{-3}$ and by
$0.1125$, in the opposite direction, at $3{\times}10^{-4}$. A second
variant therefore selects each model's rate by 20-step rollout on the
validation trajectory (17 windows), with a fixed tie-break rule
resolving FNO3d, whose two best rates are not separable there.
The validation choice agrees with the test ordering for 3/3 models on
$k_f{=}3$ and 4/4 on $Re_\lambda 70$ trained on six trajectories, no
selected cell diverges, and the four selected models score
$0.766$--$0.786$ on $Re_\lambda 70$ against persistence $0.822$.

\paragraph{Training trajectories.}
Adding training trajectories, which rank lower in the quality score
(with the normalization refit), helps Transolver on both configurations
tested: on $\tau{=}1$, three to six trajectories lower its error by
$0.014$ [$0.010$, $0.018$] (3/3, EPT $0.23\rightarrow0.40$), and on
$Re_\lambda 70$ three, six, and twelve trajectories give $0.789$,
$0.786$, and $0.701$. FNO3d stays within its seed range ($\tau{=}1$:
$+0.011$ [$-0.030$, $+0.052$]; $Re_\lambda 70$: $0.929$, $1.060^{\dagger}$,
$0.822$) while its EPT lengthens ($2.6$, $3.3$, $4.4$ frames on
$Re_\lambda 70$). At matched optimizer steps, twelve trajectories at
stride 4 against three at stride 1 lower Transolver by $0.019$
[$0.016$, $0.022$] (3/3) and FNO3d by $0.055$ [$-0.003$, $0.108$].

\paragraph{Sampling density.}
The frame stride sets the number of training pairs per epoch. Transolver
degrades as sampling thins (stride 1, 4, and 8: $0.742$, $0.810$,
$0.823$ on $k_f{=}3$; $0.719$, $0.789$, $0.806$ on $Re_\lambda 70$).
FNO3d on $Re_\lambda 70$ is non-monotone ($0.877$, $0.929$, $0.775$),
with enstrophy ratios $14.8$, $11.6$, and $0.72$. Its stride-1 and
stride-8 values differ by $0.102$, less than the spread of its five
stride-4 seeds ($0.293$), but the stride-8 value lies below the lowest
of them ($0.798$), so the three strides support no general statement about
data volume for this model.
Table~\ref{tab:app-stride1} lists every reference-seed cell trained at stride 1
and its effect on the persistence plateau discussed in
Section~\ref{sec:rollout}; early stopping makes the total number of
optimizer steps differ between runs, so these are comparisons between
recipes rather than at matched steps.

\paragraph{Unrolled training.}
We fine-tune the reference-seed weights for five epochs with a
four-step unrolled loss and compare against a single-step fine-tune with
matched optimizer steps and windows. On $k_f{=}3$ FNO3d reaches $0.669$ against $0.816$ for the
single-step fine-tune (lower on 3/3 trajectories) and TFNO
$1.73^{\dagger}$ against $150.9^{\dagger}$; on $Re_\lambda 70$ FNO3d reaches
$0.703$ against $0.950$ ($-0.247$ [$-0.266$, $-0.228$], 3/3) and TFNO
$0.818$ against $13.99^{\dagger}$ ($-13.2$ [$-14.1$, $-12.3$], 3/3), no
longer diverging. The single-step fine-tune does not improve on its
starting point ($0.810$ and $0.929$ for FNO3d).

\paragraph{Evaluation protocol.}
Holding the reference-seed checkpoint fixed and changing only the
evaluation, native full-field inference gives nRMSE $0.810$ on
$k_f{=}3$, tiling the domain into eight blocks gives $1.139$ with the
enstrophy ratio rising from $4.5$ to $30.4$ and the high-band spectral
error from $8.8$ to $84.8$, and adding a halo of eight cells recovers
only part of it ($1.071$, high-band $61.4$). Appendix~\ref{app:insights}
gives the mechanism.

\subsection{Conditioning on the forcing state}
\label{app:conditioning}
\begin{table}[H]\centering
\caption{Equations-informed integrator, 20-step nRMSE under three forcing
inputs: none (a); the current forcing state estimated from two past
frames and propagated by its OU conditional mean (b); the true forcing
(c). Share = $[(a){-}(b)]/(a)$, the part of the forcing-free error the
estimate removes.}
\label{tab:app-integrator}\footnotesize\setlength{\tabcolsep}{3.4pt}
\begin{tabular}{@{}lccccc@{}}
\toprule
Config (windows) & (a) & (b) & (c) & share \\
\midrule
$k_f{=}3$ (48) & 0.411 & 0.248 & 0.0009 & 40\% \\
$k_f{=}4$ (48) & 0.404 & 0.238 & 0.0008 & 41\% \\
$\tau{=}1$ (48) & 0.489 & 0.391 & 0.0017 & 20\% \\
$Re_\lambda 86$ (66) & 0.446 & 0.290 & 0.0011 & 35\% \\
\midrule
hot-start decay (12) & $3.4{\times}10^{-5}$ & $3.6{\times}10^{-5}$ & $2.7{\times}10^{-5}$ & --- \\
\bottomrule
\end{tabular}
\end{table}

\begin{table*}[t]\centering
\caption{History and forcing inputs under the reference recipe (25
epochs), scored on identical target frames ($Re_\lambda 86$: 131
windows; $k_f{=}4$ and $\tau{=}1$: 48; hot-start: 12). History: four
input frames (K4) against one (K1); paired 95\% CI and trajectories on
which K4 is lower for seed-0 rows, three-seed means for $k_f{=}4$
(K4 has 2--5\% fewer optimizer steps). Forcing channel: an input channel
carrying the forcing estimated from two past frames, minus the same
architecture with that channel zeroed; three-seed median [range] with
trajectories lower on each seed, unless marked one seed.}
\label{tab:app-history}\footnotesize\setlength{\tabcolsep}{4pt}
\begin{tabular}{@{}llll@{}}
\toprule
Config & Model & History: K4 $-$ K1 (20-step nRMSE) & Forcing channel $-$ zeroed channel \\
\midrule
$Re_\lambda 86$ & FNO3d & $+0.099$ [$+0.061$, $+0.131$], 0/7 & $-0.174$ [$-0.198$, $-0.119$], 7/7 on each seed \\
$Re_\lambda 86$ & Transolver & $-0.062$ [$-0.070$, $-0.045$], 6/7 & $-0.032$ [$-0.033$, $-0.029$], 7/7 on each seed \\
$k_f{=}4$ & FNO3d & K4 1.161 (seeds 0.98--1.51) vs.\ K1 0.678; step 1: 0.098 vs.\ 0.135 & $-0.065$ [$-0.074$, $-0.038$], 9/9 \\
$k_f{=}4$ & Transolver & K4 0.934 vs.\ K1 0.800; EPT 1.08 vs.\ 0.82 & $+0.002$ [$+0.001$, $+0.002$], one seed \\
$\tau{=}1$ & FNO3d & $+0.030$ [$-0.011$, $+0.071$], 1/3 & $-0.095$ [$-0.120$, $-0.070$], one seed \\
$\tau{=}1$ & Transolver & $+0.276$ [$+0.222$, $+0.331$], 0/3 & $-0.015$ [$-0.017$, $-0.012$], one seed \\
hot-start & FNO3d & $-0.044$ [$-0.046$, $-0.042$], 3/3 & --- \\
hot-start & Transolver & $-0.069$ [$-0.076$, $-0.062$], 3/3 & --- \\
\bottomrule
\end{tabular}
\end{table*}

A single frame omits the forcing state, which the OU drive carries from
frame to frame. Table~\ref{tab:app-integrator} bounds the effect for the
equations-informed integrator. (a)$-$(b) is the error that the explicit
past-frame estimate of the current forcing removes; (b)$-$(c) is the
remainder, which includes the estimation error and is therefore an
upper bound on the part no input can remove. On the hot-start decay all
three inputs agree below $10^{-4}$, so the estimator adds no error when
no forcing is present.

Table~\ref{tab:app-history} tests the two corresponding inputs for
learned operators. Without forcing the velocity field is the full
state, so history should add nothing; with OU forcing the state is
$(u,f)$ and history can partly identify $f$. On forced $Re_\lambda 86$
four frames help Transolver at every step (6/7 trajectories) and hurt
FNO3d (0/7) except in the forced band $|k|<k_f$ (7/7). On $k_f{=}4$ both
models improve at step 1 and worsen over 20 steps, and the 20-step mean
and EPT can order K1 and K4 oppositely, so both are reported. On the
decay control both models also improve (3/3), although the single frame
is the full state there, so part of the history gain is unrelated to
forcing, and we read only the forced-minus-decay difference within one
model and band as forcing-related. The forcing channel lowers error
against its zeroed counterpart on 3/3 seeds in all three
model--configuration pairs trained with three seeds. For FNO3d the effect is within the zeroed-channel cross-seed
spread ($0.17$ vs.\ $0.26$ and $0.065$ vs.\ $0.066$), and for Transolver
it is about five times the spread ($0.032$ vs.\ $0.006$). Against the
plain single-frame baseline the forcing-channel FNO3d is not better
(three-seed median $+0.003$ [$-0.213$, $+0.049$]) and Transolver is
($-0.032$ [$-0.038$, $-0.031$]). Because the estimate uses two frames,
these are upper bounds on what the current forcing state contributes.

\paragraph{Forced-to-decay transfer.}
The transfer changes frame spacing and statistics as well as forcing,
and no control isolates forcing alone. Stepping the integrator at the
forced frame spacing on hot-start alone costs $0.340$ [$0.328$,
$0.352$] (persistence $1.019$), the same order as the forcing-free error
on the forced configurations. Giving the forcing-channel FNO3d the true
decay forcing (zero) makes it better than its zeroed counterpart
($-0.066$ [$-0.100$, $-0.034$], 3/3); against an ordinary single-frame
transfer it is $-0.004$ [$-0.034$, $+0.026$] under the source model's
normalization and $-0.063$ [$-0.104$, $-0.025$] under the target's,
while the corresponding Transolver worsens and diverges. A forced-trained
FNO3d with $k_f{=}4$ as source reaches $0.992$ on hot-start against
persistence $1.017$ (not separable under the target normalization,
$-0.025$ [$-0.094$, $+0.045$]; $1.148$ under its source normalization)
and $0.831$ on Saffman against $0.899$, below persistence under both
normalizations. The hot-start test windows cover the first $5.4$ time
units after the drive is removed, and this model's error is highest in
the earliest of them.

\subsection{Training coverage}
\label{app:joint}
\begin{table*}[t]\centering
\caption{Joint training across regimes (seed 0, reference recipe):
paired difference in 20-step nRMSE on the target test set, joint model
minus the reference model named in the third column, with 95\% CI and
the trajectories on which the joint model is lower; target / source
normalization where both were scored. Step-matched controls train the
single source for the same number of optimizer steps as the joint model.}
\label{tab:app-joint}\footnotesize\setlength{\tabcolsep}{4pt}
\begin{tabular}{@{}lllc@{}}
\toprule
Axis (target) & Model & Reference & Joint $-$ reference \\
\midrule
$Re_\lambda\{55,70\}\rightarrow 86$ & Transolver & $Re_\lambda 70$ single source & $-0.0008$ [$-0.0016$, $+0.00003$] \\
 & FNO3d & $Re_\lambda 70$ single source & $+0.134$ [$+0.110$, $+0.159$], 0/7 \\
$\{k_f{=}3$, rot.\ moderate$\}\rightarrow$ rot.\ strong & Transolver & step-matched $k_f{=}3$ control & $-0.020$ [$-0.025$, $-0.015$] / $-0.046$ [$-0.049$, $-0.042$], 3/3 \\
 & Transolver & reference $k_f{=}3$ source & $+0.024$ [$+0.018$, $+0.030$] / $+0.009$ [$+0.008$, $+0.011$], 0/3 \\
 & FNO3d & reference $k_f{=}3$ source & $-0.034$ [$-0.138$, $+0.070$] / $-0.119$ [$-0.163$, $-0.076$] \\
$\{Re_\lambda 86$, hot-start$\}\rightarrow$ hot-start & Transolver & step-matched $Re_\lambda 86$ control & $+0.0003$ [$-0.0013$, $+0.0019$] \\
 & FNO3d & step-matched $Re_\lambda 86$ control & $+0.021$ [$-0.005$, $+0.047$] \\
\bottomrule
\end{tabular}
\end{table*}

Table~\ref{tab:app-joint} trains one model on two regimes and scores it
on a target regime, against the reference single-source model and, where
the joint model sees more optimizer steps, a single-source control
trained for the same number of steps. On none of the three axes is the
joint model better than both references under both normalizations. On
the rotation axis the Transolver joint model beats the step-matched
control but not the reference single-source model, so part of its
apparent gain comes from the control degrading with longer training. For
FNO3d, early stopping ends the step-matched control before the joint
model's step count, leaving only the comparison with the reference
source, which is separable under one normalization and not the
other. Adding hot-start data to flagship training improves Transolver on
hot-start only through the extra optimizer steps ($-0.135$ [$-0.148$,
$-0.122$] for the step effect, $+0.0003$ for the data at matched steps).

\subsection{Split and slice robustness}
\label{app:splits}
\begin{table}[H]\centering
\caption{Split robustness (seed 0, reference recipe): 20-step nRMSE on
each split's own test set, with the trajectories on which the model is
lower than persistence. Every split refits the input normalization.
$Re_\lambda 86$: all windows of seven test trajectories; $k_f{=}4$: 51
windows of three ($^{\ast}$48 aligned windows).}
\label{tab:app-splits}\footnotesize\setlength{\tabcolsep}{3pt}
\begin{tabular}{@{}llccc@{}}
\toprule
Config & Split & FNO3d & Transolver & persist. \\
\midrule
$Re_\lambda 86$ & quality-ranked & 1.022 (0/7) & 0.823 (7/7) & 0.860 \\
 & random & 1.038 (0/7) & 0.824 (7/7) & 0.865 \\
 & anti-ranked & 1.023 (0/7) & 0.816 (7/7) & 0.863 \\
 & random (second) & 1.007 (0/7) & 0.829 (7/7) & 0.860 \\
\midrule
$k_f{=}4$ & quality-ranked & 0.661$^{\ast}$ & 0.800$^{\ast}$ & 0.822 \\
 & random & 0.659 (3/3) & 0.801 (3/3) & 0.822 \\
 & anti-ranked & 0.667 (3/3) & 0.799 (3/3) & 0.820 \\
\bottomrule
\end{tabular}
\end{table}

Table~\ref{tab:app-splits} compares the quality-ranked split with two
random splits and an anti-ranked split that trains on the three
lowest-ranked trajectories. On $Re_\lambda 86$ the four splits span
$0.030$ for FNO3d and $0.013$ for Transolver, and under every split
FNO3d is better than persistence on 0/7 trajectories and Transolver on
7/7. We compare the split range with the seed range in the same cell, both on full trajectories: FNO3d $0.030 <
0.103$ and Transolver $0.013 > 0.004$, so the split effect is smaller
than the seed effect for FNO3d and larger for Transolver without changing
any rank. $k_f{=}4$ shows the same pattern (FNO3d $0.008 < 0.047$,
Transolver $0.003 > 0.001$). On the test trajectories that all
$Re_\lambda 86$ splits share, FNO3d trained on the alternative splits is
higher by $0.035$--$0.048$ (each separable) and Transolver lies within
$-0.008$ to $+0.006$.

\section{Compute and Reproducibility}
\label{app:compute}
Corpus production required approximately 105 GPU-hours: $872\,T_L$ of
sampled evolution across the 134 released trajectories plus spin-up,
extrapolated from a measured $101.5$\,ms per step at $256^3$ in fp64 on an
RTX~5090. Wall-clock time was not recorded per configuration, and the assumption
of $30\,T_L$ of spin-up per trajectory is the dominant uncertainty, which
places the corpus total in the range 80--130 GPU-hours. The reference
benchmark required approximately 97 GPU-hours: 166 training-plus-evaluation
runs of the learned baselines across three workstations, with median
per-run wall-clock in Table~\ref{tab:cost}, plus the non-learning
references, which are evaluation only. Together this is roughly 200
GPU-hours on consumer GPUs. The flagship configuration is independently
reproduced across hardware and library versions with consistent
statistics.
A main-table cell is reproduced in four steps. The benchmark slice
written by \texttt{make\_benchmark\_slice.py} fixes the training,
validation, and test seeds of each configuration;
\texttt{train\_benchmark.py} trains one model on one seed under the
reference recipe (frame stride 4, 25 epochs, learning rate $10^{-3}$);
\texttt{eval\_benchmark.py} scores the selected checkpoint on the test
windows (one every 8 frames, first 51) and writes one result row per run,
which records the recipe it was trained with; the rows of all seeds of a
cell give its median, range, and divergence count. The decision rules
(divergence thresholds, window cap, checkpoint selection, learning-rate
tie-break, and the separability test) were set before any test-set
evaluation and apply uniformly to all runs.
Bit-exact trajectory regeneration is hardware-bound (GPU RNG); statistical
reproduction (spectra, acceptance verdicts) is hardware-independent from
the released solver, configuration files, and seeds. The acceptance
scripts, benchmark metrics, and solver each carry their own audited test
suites.

\begin{table}[t]
\centering
\caption{Per-run cost on one RTX~5090: median wall-clock (training +
evaluation) over $n$ measured runs, and peak training memory.}
\label{tab:cost}
\footnotesize
\setlength{\tabcolsep}{5pt}
\begin{tabular}{@{}lcccc@{}}
\toprule
\textbf{Model} & \textbf{Params} & \textbf{Peak mem.} & \textbf{Wall-clock/run} & $n$ \\
\midrule
FNO3d          & 28.3\,M & 5.0\,GB  & 0.56\,h & 36 \\
TFNO     & 30.3\,M & \textit{checkpointed} & 0.66\,h & 15 \\
Transolver     & 29.7\,M & \textbf{2.8\,GB}  & \textbf{0.21\,h} & 18 \\
Spectral U-Net & 28.2\,M & 6.8\,GB  & 0.85\,h & 12 \\
DeepONet-3D    & 7.0\,M  & 24.5\,GB & 0.28\,h & 3 \\
\bottomrule
\end{tabular}
\\[2pt]
{\footnotesize TFNO peak memory is set by its gradient-checkpointing
schedule. Medians pool all tasks; forecasting runs are the longest
(FNO3d $1.25$--$1.95$\,h, TFNO $1.5$--$2.25$\,h).}
\end{table}

\end{document}